\documentclass[letterpaper,fleqn,usenatbib]{mnras}

\usepackage{newtxtext,newtxmath}

\usepackage[T1]{fontenc}
\usepackage{ae,aecompl}

\usepackage{tabularx}
\usepackage[normalem]{ulem}
\usepackage{multirow}
\usepackage{upgreek}
\usepackage{graphicx}	
\usepackage{amsmath}	
\usepackage{amssymb}	
\usepackage[multiple]{footmisc} 

\newcommand{\RN}[1]{%
	\textup{\uppercase\expandafter{\romannumeral#1}}%
}

\title[Radio eclipses of spider pulsars]{Study of spider pulsar binary eclipses and discovery of an eclipse mechanism transition}

\author[E. J. Polzin et al.]{
E. J. Polzin,$^{1}$\thanks{E-mail: elliott.polzin@manchester.ac.uk (EJP)}
R. P. Breton,$^{1}$
B. Bhattacharyya,$^{2}$
D. Scholte,$^{1}$
C. Sobey,$^{3}$
\newauthor
B. W. Stappers$^{1}$
\\
$^{1}$Jodrell Bank Centre for Astrophysics, Department of Physics and Astronomy, The University of Manchester, Manchester M13 9PL, UK\\
$^{2}$National Centre for Radio Astrophysics, Tata Institute of Fundamental Research, Pune University, Pune 411007, India\\
$^{3}$CSIRO Astronomy and Space Science, PO Box 1130 Bentley, WA 6102, Australia\\
}

\date{Accepted XXX. Received YYY; in original form ZZZ}

\pubyear{2020}

\begin{document}
\label{firstpage}
\pagerange{\pageref{firstpage}--\pageref{lastpage}}
\maketitle

\begin{abstract}
We present a comparative study of the low-frequency eclipses of spider (compact, irradiating binary) PSRs B1957+20 and J1816+4510. Combining these data with those of three other eclipsing systems we study the frequency dependence of the eclipse duration. PSRs B1957+20 and J1816+4510 have similar orbital properties, but the companions to the pulsars have masses that differ by an order of magnitude. A dedicated campaign to simultaneously observe the pulsed and imaged continuum flux densities throughout the eclipses reveals many similarities between the excess material within the two binaries, irrespective of the companion star properties. The observations show that the pulsar fluxes are removed from the line of sight throughout the main body of the eclipses. For PSR J1816+4510 we present the first direct evidence of an eclipse mechanism that transitions from one that removes the pulsar flux from the line of sight to one that merely smears out pulsations, and claim that this is a consequence of scattering in a tail of material flowing behind the companion. Inferred mass loss rates from the companion stars are found to be $\dot{M}_{\text{C}} \sim 10^{-12}~M_\odot$~yr$^{-1}$ and $\dot{M}_{\text{C}} \sim 2 \times 10^{-13}~M_\odot$~yr$^{-1}$ for PSR B1957+20 and PSR J1816+4510, respectively; seemingly too low to evaporate the stars within Hubble time. Measurements of eclipse durations over a wide range of radio-frequencies show a significant dependence of eclipse duration on frequency for all pulsars, with wider eclipses at lower-frequencies. These results provide a marked improvement in the observational constraints available for theoretical studies of the eclipse mechanisms.
\end{abstract}




\section{Introduction}\label{sec: intro}
\begin{figure}
	\centering
	\includegraphics[width=0.6\columnwidth]{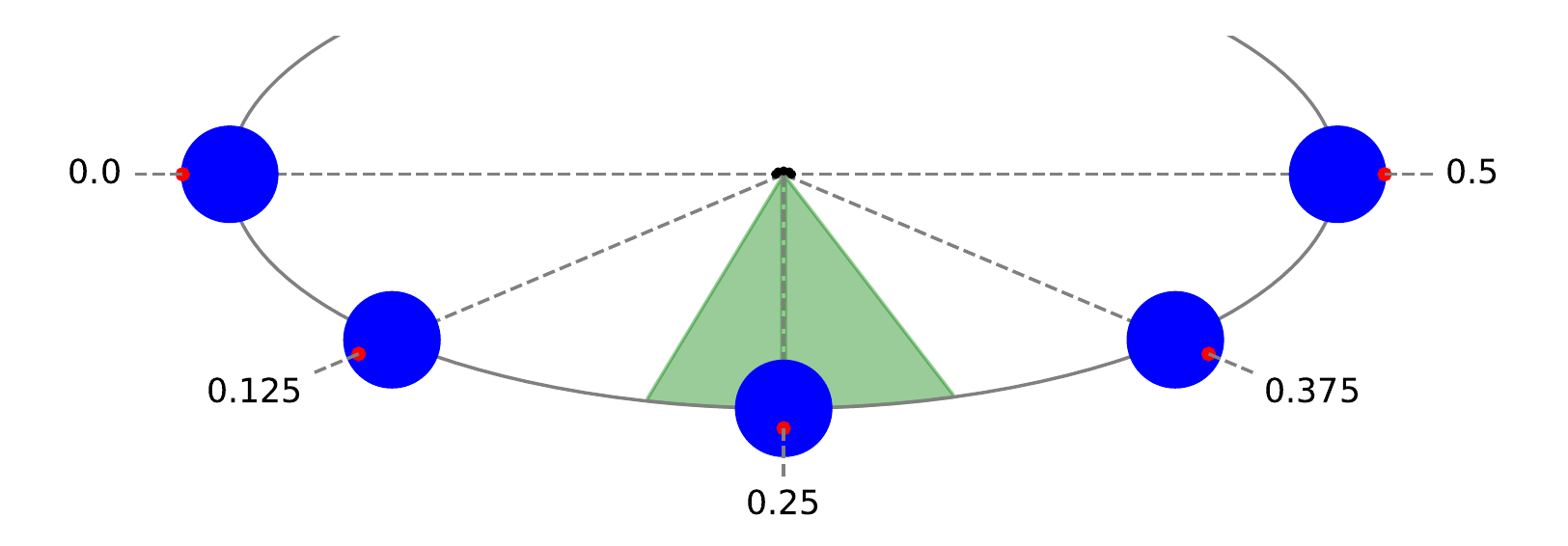}
	\caption[Schematic of the projection of the PSR B1957+20 system on the sky]{Expected projection of the PSR B1957+20 system on the sky, assuming an inclination angle of $65^{\circ}$ \citep{vbk11}. The schematic shows 5 snapshots of the system at the labelled orbital phases, with the central, small black dot representing the pulsar, blue circles representing the expected size of the companion star \citep[$R_C \approx 0.25\,R_{\odot}$;][]{vbk11}, and small red dots marking the centre point of the night-side of the companion star (i.e. the furthest point from the pulsar). The companion star and orbit are approximately to scale assuming the radio timing orbital parameters of \citet{aft94}, and pulsar and companion masses of $2.4\,M_{\odot}$ and $0.035\,M_{\odot}$, respectively. The companion passes closest to our line-of-sight towards the pulsar -- directed out of the page -- at an orbital phase of 0.25. The shaded green segment of the orbit represents the typical eclipse region at 149\,MHz, as presented in this paper.}
	\label{fig:B1957_geometry}
\end{figure}

Spider pulsar systems are characterised by having a low-mass companion star in a compact orbit with an energetic millisecond pulsar (MSP) resulting in heavy irradiation of the companion by the pulsar's wind. The spider pulsar population has been observed to have a clearly bi-modal distribution of companion star masses \citep{r11,ssc+19} made up of two distinct sub-groups: black widows (BW) with companion star masses $\sim0.01$--$0.05 M_\odot$, and redbacks (RB) with companion star masses $\sim0.1$--$1 M_\odot$. A large proportion of the spiders, whether BWs or RBs, have been observed to exhibit (quasi-)periodic eclipses of the pulsars' radio emission \citep[e.g.][]{fst88,lmd+90} that are generally attributed to excess material in the orbits -- that has been driven from the companion stars by the pulsar irradiation \citep{p91,vv88,peb+88,krs+88} -- interfering with the propagation of the radio emission. Studies of such eclipses are key for understanding mass loss from the irradiated companion stars, the properties of the medium causing the eclipses, interactions between the pulsar wind and the eclipse medium, and the mechanisms responsible for the apparent attenuation of pulsar radio emission during the eclipse. In the years after the initial BW discovery \citep{fst88} there were a number of excellent early works \citep[e.g.][]{rt91,sbl+01} investigating the observed radio eclipses. However, unfortunately, a lack of further in-depth eclipse analyses -- largely as a result of difficult observing requirements and (until recently) a low number of known spider pulsars -- has meant slow progress in reaching an understanding in any of these topics. However, the last few years have marked a revival of the field with detailed and novel studies beginning to give important insight into the nature of eclipsing pulsar systems \citep[e.g.][]{bfb+16,myc+18,llm+19}.\\
Observing spider pulsars at low-radio frequencies offers a particularly good opportunity to further our understanding of the eclipses. There are several reasons for this: firstly, it is observed that the radio emission from pulsars is generally significantly brighter toward lower-frequencies, with average spectral indices of $-1.8$ \citep{mkk+00} or $-1.4$ \citep{blv13}. Secondly, the Low-Frequency Array \citep[LOFAR;][]{v+13} offers unprecedented sensitivity at frequencies $< 200$\,MHz, allowing high signal-to-noise detections. Finally, the observable effects of scattering and dispersion scale strongly with the inverse of frequency -- $\nu^{-4}$ and $\nu^{-2}$, respectively -- and when twinned with the wide fractional bandwidth of LOFAR allow precise measurements of increased dispersion measure (DM) and pulse scattering timescale ($\tau$) at eclipse edges. These increases are expected as the eclipse material begins to affect the propagation of the radio emission. An additional advantage provided by LOFAR is the capability to simultaneously record high-time resolution beamformed data and also visibilities that can be used to produce images of the sky. This capability allows direct discrimination between eclipse mechanisms that remove pulsar flux from the line of sight, from those that only smear out the pulsations. The success of such observations for determining many of the useful eclipse metrics was demonstrated in \citet{pbc+18} (hereafter P18) for PSR J1810+1744, inspiring a programme of similar observations for other spider pulsars.\\
Two spider systems in particular were chosen for this study, namely, PSR J1816+4510 and PSR B1957+20 (hereafter J1816 and B1957, respectively), both of which were known to be relatively bright at low-frequency and easily detectable with LOFAR \citep{kvh+16}. Furthermore, both systems have similar orbital periods, with comparable pulsar--companion star separations, while containing companion stars with masses separated by nearly an order of magnitude \citep{aft94,vbk11,slr+14,kbv+13}, providing the opportunity to investigate how this affects the eclipses. Table~\ref{Table: psr_params} lists previously estimated parameters of these two binaries that are most pertinent to the following eclipse analyses.\\
B1957 was the first BW to be discovered \citep{fst88}, and is one of the most intensively studied. Optical observations revealed the inner face of the tidally locked companion star to be strongly irradiated \citep{kdf88,vac+88}, and a subsequent study of the companion's radial velocity suggested that the system harbours a massive pulsar -- $M_{\text{PSR}} = (2.40 \pm 0.12) M_{\odot}$ -- in a 9.2\,h orbit inclined at $65^{\circ}$ \citep{vbk11}. In the radio domain, regular eclipses have been observed in a range of frequencies over $300$\,MHz \citep{fbb+90,rt91}. Based on these observations, there have been attempts to model both the outflow from companion star \citep{tb91} and the mechanisms behind the eclipses \citep[e.g.][]{t+94}, however these have remained largely inconclusive. In Fig.~\ref{fig:B1957_geometry} we show a schematic of the expected orbit of B1957 as viewed in the sky.\\
\noindent
J1816, on the other hand, was only discovered much more recently, with its radio eclipses and pulse arrival delays leading to it being classified as a RB \citep{slr+14}. However, follow-up studies in the optical domain revealed the companion star to have unique properties among the other known RBs, with an unusually high temperature and metallicity best resembling a proto-white dwarf \citep{kbv+13}, which has since brought this classification into doubt \citep{itl+14}. Although not a `stereotypical' RB -- a somewhat vaguely defined term in this context as the number of well characterised RBs is low \citep[c.f.][]{ssc+19} -- it still represents the class in that the tidally locked, irradiated companion is roughly an order of magnitude more massive than that in BWs. Interestingly, J1816 also appears to harbour a massive pulsar -- $M_{\text{PSR}} \sin^3i = (1.84 \pm 0.11) M_{\odot}$ -- but the inclination, $i$, of the 8.7\,h orbit is relatively unconstrained \citep{kbv+13}, and as such its projection on the sky is not known.\\
Measuring radio eclipse durations as a function of observing wavelength, in an individual spider pulsar system, has the potential to discriminate between eclipse mechanisms, conditional on the assumptions made about the electron density and temperature distributions and magnetic properties of the eclipse medium. Alternatively, by assuming a given eclipse mechanism, certain properties of the eclipse medium can be inferred through the frequency dependence of the observed eclipses. In a thorough critique of possible eclipse mechanisms for B1957 and PSR B1744$-$24A (Ter5A), \citet{t+94} directly refer to the observed degree of frequency dependence of the eclipse durations as being able to provide evidence against a refractive mechanism. Here the authors showed that it is not plausible for both the small measured pulse arrival time delays at eclipse edges, and the relatively strong sensitivity of eclipse duration on frequency, to be simultaneously observed for a refractive mechanism with realistic density distributions for the eclipse material. Further, both \citet{t+94} and \citet{kmg00} cite the consistency of the predicted frequency dependence from cyclotron-synchrotron absorption with those observed in B1957 as evidence in favour of such a mechanism being responsible for the eclipses.\\
However, there are significant problems associated with the reliable measurement of eclipse durations as a function of frequency. Generally, spider pulsar eclipse durations are of the order of 1\,h, and the change in duration with frequency tends to be of the order of a few mins for typical observing bandwidths. As such, long duration observations with large fractional frequency coverage are required in order to make precise measurements, which can be difficult to achieve with limited availability of telescope time. Furthermore, the very definition of eclipse duration is not robust, and depends on the signal-to-noise detection of the pulsar away from the eclipse. As the disappearance and re-emergence of flux at eclipse edges occurs over a finite time, a higher signal-to-noise observation of a pulsar, at a given frequency, will probe deeper into the eclipse region than an equivalent observation with lower signal-to-noise. A third problem arises from the time-variable nature of the eclipses; due to the necessity for wide-frequency coverage, observations must often be performed with different telescope facilities, which can lead to large temporal separations between measurements of the eclipse durations at different frequencies, in which time the eclipse medium may have significantly changed. \citet{sbl+01} highlight the latter in their attempt to measure the frequency dependence for PSR J2051$-$0827, whereby the variability in eclipse durations at each individual frequency completely masks any genuine dependence on frequency. The collective effect of these issues has been to allow measurements of significant frequency dependencies of eclipse durations for only a few pulsar systems \citep{ntt+90,rt91,bfb+16} -- at least to our knowledge, some may have been missed due to only sporadic reporting of these measurements over the last $\sim30$~yrs.\\
To vastly increase the observational samples available for future studies, we accompany our observations of the frequency dependence of the J1816 and B1957 eclipses with those from a further set of spider pulsars: PSR J1810+1744 -- a BW in a 3.6\,h orbit \citep{hrm+11}, PSR J2051$-$0827 -- a BW in a tight, 2.4\,h orbit \citep{s+96} and PSR J2215+5135 -- a RB pulsar in a 4.2\,h orbit \citep{hrm+11} that is thought to not only harbour a particularly massive pulsar \citep[$2.27^{+0.17}_{-0.15} M_{\odot}$;][]{lsc18} but is also suspected to potentially undergo transitions between accretion-powered X-ray states and its current radio pulsar state due to its observed high X-ray luminosity, similar to those seen in the known transitioning low-mass X-ray binary -- RB pulsars \citep{l14}. Where possible we incorporate previously published eclipse duration data for the same sources. In Table~\ref{Table: psr_params} we summarise the relevant parameters for each of the binary systems in this study.\\
The work in this paper is presented as follows: Section~\ref{sec: obs} briefly introduces the observations used, Section~\ref{sec: analysis} then details the analysis methods applied to the observational data, Section~\ref{sec: ecl_obs} presents the eclipses of J1816 and B1957, Section~\ref{sec: freq_dependence} presents measured eclipse durations as a function of radio-frequency for these two sources along with a number of other spider pulsars and Section~\ref{sec: discuss} then discusses the eclipses of J1816 and B1957 and considers eclipse behaviour in the context of spider pulsars in general. Finally, the findings are concluded in Section~\ref{sec: conclusion}.

\section{Observations}\label{sec: obs}
A dedicated campaign was undertaken to observe J1816 and B1957 using the LOFAR Core High Band Antennas (HBA; 110--188\,MHz) in the commensal beamformed and imaging mode. This made use of the COBALT correlator \citep{bmn+18} to simultaneously record complex-voltage data and correlated raw visibility data. The complex-voltages were recorded with spectral and temporal resolutions of 195.3\,kHz and 5.12\,$\mu$s, respectively, while the visibilities were recorded with resolutions of $\sim3$\,kHz and 2\,s. The $\sim2$\,km maximum baseline length of the LOFAR Core stations resulted in a single tied-array beam with a width of $\sim165$\,arcsec at 150\,MHz with uniform station weighting.\\
The observing strategy was to observe 4 separate eclipses for each of J1816 and B1957. The observation dates were selected such that, for each pulsar, 2 eclipses were observed in close succession (within a few days), followed by a gap of a few months before observing two further eclipses in close succession, as shown in Table~\ref{Table: obs_ecl}. This observing schedule was chosen with the aim of investigating the eclipses on different timescales. In addition, the schedule was designed so that the eclipse of each source was centred near the meridian, hence minimising the effects of telescope sensitivity reductions that can occur for low-elevation beam pointings due to shadowing of the HBA tiles. Due to the relatively long orbital periods of the two sources, the observations were chosen to be 4.5\,h in duration, corresponding to approximately half an orbit, ensuring sufficient coverage of out-of-eclipse phases to allow measurements of the baseline DMs, flux densities and timing parameters. Complementary to this dedicated project, in 2014-Oct we had already obtained a single full-orbit observation of J1816 utilising the same simultaneous interferometric and beamformed mode of the LOFAR Core, thus further aiding in the investigation of time-variable eclipse phenomena.\\
\begin{table*}
	\centering
	\caption[Observations for eclipses of PSRs J1816+4510 and B1957+20]{List of observations of PSR J1816+4510 and PSR B1957+20, sorted by date of observation, used to characterise the low-frequency eclipses. $^{\text{a}}$~BF -- beamformed, Im -- image-plane}
	\label{Table: obs_ecl}
	\begin{tabular}{lccccc}
		\hline	
		PSR & MJD & Telescope & Centre Frequency (MHz) & Duration (h) & Type$^{\text{a}}$ \\
		\hline
		J1816+4510 & 56947.546 & LOFAR & 149 & 10 & BF+Im \\
		B1957+20 & 58333.841 & LOFAR & 149 & 4.5 & BF+Im \\
		B1957+20 & 58336.883 & LOFAR & 149 & 4.5 & BF+Im \\
		J1816+4510 & 58344.549 & uGMRT & 650 & 6.5 & BF \\
		J1816+4510 & 58358.722 & LOFAR & 149 & 4.5 & BF+Im \\
		J1816+4510 & 58363.741 & LOFAR & 149 & 4.5 & BF+Im \\
		J1816+4510 & 58480.379 & LOFAR & 149 & 4.5 & BF+Im \\
		J1816+4510 & 58484.369 & LOFAR & 149 & 4.5 & BF+Im \\
		B1957+20 & 58490.823 & LOFAR & 149 & 4.5 & BF+Im \\
		B1957+20 & 58492.408 & LOFAR & 149 & 4.5 & BF+Im \\
		\hline
	\end{tabular}
\end{table*}\noindent
Many thorough studies of the eclipse phenomena of B1957 exist for observations at frequencies above 300\,MHz \citep[e.g.][]{fbb+90,rt91} that complement our LOFAR observations, hence these results are used in our interpretation of the low-frequency data. On the other hand, no such in-depth studies exist for J1816 -- the eclipse and corresponding TOA delays are reported in \citet{slr+14}, but not deeply investigated -- thus we obtained a dedicated eclipse observation at higher-frequency to complement the LOFAR data. For this, we utilised the upgraded Giant Metrewave Radio Telescope \citep[uGMRT;][]{gak+17} to make a 6.5\,h observation of J1816 with all available antennas in a 550--750\,MHz band, pausing every $\sim1.5$\,h to phase calibrate the array for $\sim10$\,mins. The raw data were recorded with spectral and temporal resolutions of 48.8\,kHz and 81.92\,$\mu$s, respectively.\\
Further observations were made of three other spider pulsars utilising multiple telescopes to get wide frequency coverage. Specifically, observations were made of PSRs J1810+1744, J2051$-$0827 and J2215+5134 with combinations of LOFAR, uGMRT and the Parkes Telescope, as listed in Table~\ref{Table: obs_freq}. For a number of these sources there are previously published eclipse measurements which we include in our analysis and these are cited in Table~\ref{Table: obs_freq} and the relevant sections.\\
\begin{table*}
	\centering
	\caption[Observations used in frequency dependence study]{List of observation dates and bandwidths used to study the frequency dependence of eclipses. $^{\text{a}}$ \citet{pbc+18}; $^{\text{b}}$ \citet{rt91}; $^{\text{c}}$ Robert Main priv. comm.; $^{\text{d}}$ \citet{pbs+19}; $^{\text{e}}$ \citet{sbl+01}; $^{\text{f}}$ \citet{bfb+16}.}
	\label{Table: obs_freq}
	\begin{tabular}{lcccccc}
		\hline	
		PSR & LOFAR & WSRT & uGMRT & Parkes Telescope & William E. Gordon & Lovell Telescope \\
		&       &      &       &                  & Telescope & \\
		&       &      &       &                  & (Arecibo) & \\
		\hline
		J1810 & 110--188\,MHz & 310--381\,MHz & 300--500\,MHz, & & & \\
		+1744 & 2012-Dec--2015-Feb$^{\text{a}}$ & 2011-Jun$^{\text{a}}$ & 650--850\,MHz & & & \\
		& & & 2018-Aug & & & \\
		&  &  &  &  &  & \\
		J1816 & 111--186\,MHz &  & 550--750\,MHz & & & \\
		+4510 & 2018-Aug--2019-Jan & & 2018-Aug & & & \\
		&  &  &  &  &  & \\
		B1957 & 111--186\,MHz &  &  &  & 318--606\,MHz & \\
		+20 & 2018-Aug--2019-Jan & & & & 1988-Mar--1990-Dec$^{\text{b}}$ & \\
		& & & & & 2014-Jun$^{\text{c}}$ & \\
		& & & & & 2018-Apr--2018-Jul$^{\text{c}}$ & \\
		&  &  &  &  &  & \\
		J2051 & 110--188\,MHz &  & 300--500\,MHz & 705--4032\,MHz & & 234--610\,MHz \\
		$-$0827 & 2018-Dec & & 2018-Dec$^{\text{d}}$ & 2018-Dec$^{\text{d}}$ & & 1994-May--1996-Mar$^{\text{e}}$ \\
		& & & & 430--660\,MHz & & \\
		& & & & 1994-May--1997-Sep$^{\text{e}}$ & & \\
		&  &  &  &  &  & \\
		J2215 & 50--190\,MHz &  & 550--750\,MHz, & & & \\
		+5135 & 2013-Feb--2014-Jan$^{\text{f}}$ & & 1260--1460\,MHz & & & \\
		& & & 2017-Aug & & & \\
		\hline
	\end{tabular}
\end{table*}\noindent
The low-frequency data for PSR J1810+1744 from P18, consisting of multiple LOFAR observations at 149\,MHz and a single Westerbork Synthesis Radio Telescope \citep[WSRT;][]{bh74} observation at 345\,MHz, have been expanded on with a recently acquired simultaneous dual-frequency uGMRT observation, with half of the available antennas observing in a 200\,MHz band centred at 400\,MHz, and the other half in a 200\,MHz band centred at 750\,MHz. This observation was 5\,h in duration, catching two full consecutive eclipses, with short $\sim10$\,min breaks occurring every 1.5\,h in order to re-calibrate the phases between separate antennas. Total intensity data were recorded with a sampling time of 81.92\,$\mu$s and spectral resolution of 48.8\,kHz.\\
As previously noted in \citet{sbl+01}, the temporal variations in the eclipses of PSR J2051$-$0827 are of a large enough magnitude to mask any clear frequency dependence. As such, we undertook a dedicated campaign to observe the pulsar's eclipses over an extremely wide frequency range, within just a 2\,week period -- shorter than the timescales of significant eclipse variability reported in \citet{pbs+19} (hereafter P19). These observations consisted of a 2\,h LOFAR observation covering 110--188\,MHz, a 2\,h uGMRT observation covering 300--500\,MHz and a 3\,h Parkes Telescope observation covering 705--4032\,MHz utilising the new ultra-wide-bandwidth low-frequency receiver \citep{hmd+19}. Note however that eclipses in this pulsar usually only occur for observing frequencies below $\sim 1$\,GHz \citep[P19;][]{sbl+01}. The specific details of these observations are given in P19, for which they were used to study the eclipse variability of PSR J2051$-$0827.\\
Finally, we obtained a single, simultaneous dual-frequency observation of the RB pulsar PSR J2215+5135. This was a 5\,h observation, covering more than a full orbit, using the uGMRT with half of the antennas recording a 200\,MHz band centred at 650\,MHz and the other half recording a 200\,MHz band centred at 1360\,MHz. The data were recorded with a sampling time of 40.96\,$\mu$s, spectral resolution of 97.6\,kHz and $\sim10$\,min breaks occurred every 1.5\,h for phase calibration of the array.

\section{Analysis}\label{sec: analysis}
\subsection{Beamformed data}\label{sec: analysis_bf}
The complex-voltage LOFAR data for J1816 and B1957 were coherently dedispersed, folded and cleaned as part of the LOFAR Known Pulsar Pipeline \citep[PulP; ][]{ahm+10,s+11}. The resulting folded data consisted of 20--30\,s duration sub-integrations, 384 frequency channels and 512 (256) pulse phase bins for J1816 (B1957), with sub-integration duration chosen based on signal-to-noise. As the observations were made with only high-elevation beam pointings and there was no need to attain absolute flux density measurements, we did not perform thorough flux calibration. Similarly, as both sources have only small percentages of polarised flux, no polarisation calibration was performed. The wandering of orbital parameters that commonly occurs in spider pulsars \citep[e.g.][]{svf+16} was corrected for in J1816 by re-fitting the timing model available in the ATNF Pulsar Catalogue\footnote{http://www.atnf.csiro.au/people/pulsar/psrcat/} \citep{mhth05} to our new out-of-eclipse TOAs using \textsc{tempo2}\footnote{https://sourceforge.net/projects/tempo2/} \citep{hem06}, with the spin frequency derivative, F1, orbital period, PB, and time of ascending node, TASC, as free parameters. This task was much more problematic for B1957, which is known to exhibit large variations in its orbit \citep{aft94}, as the readily available timing solutions had become insufficient as a baseline model to allow for the relatively small number of TOAs from these observations to accurately correct the parameters. As such, we instead used a baseline model developed using TOAs obtained from GLOW (a network of LOFAR single stations in Germany) observations spanning 2014-Jan -- 2019-Jan (Julian Donner priv. comm.). This was optimised for our set of observations by fitting to TOAs using \textsc{tempo2}, with the epoch of periastron, T0, and projected semi-major axis of the orbit, A1, as free parameters. In a similar vein, the DM of each observation was adjusted by applying the \textsc{pdmp} tool of \textsc{psrchive}\footnote{https://psrchive.sourceforge.net/} \citep{hvm04,vdo12} to the out-of-eclipse data to obtain the `optimum' value -- that which yielded the maximum pulse signal-to-noise when integrated over the bandwidth -- and installing this with \textsc{psrchive}'s \textsc{pam} tool.\\
The higher-frequency raw uGMRT data were initially converted to filterbank format with the \textsc{sigproc}\footnote{http://sigproc.sourceforge.net/} package's \textsc{filterbank} tool, then were folded into 10\,s sub-integrations, 4096 frequency channels and 512 pulse phase bins for J1816; 20\,s sub-integrations, 4096 channels and 256 bins for PSR J1810+1744 and 60\,s (650\,MHz) or 4\,min (1360\,MHz) sub-integrations, 2048 channels and 256 bins for PSR J2215+5135, ensuring high enough signal-to-noise for eclipse analyses. For J1816 the DM and timing ephemeris optimised with the LOFAR data observed within the same month were used for folding and incoherent dedispersion. For PSRs J1810+1744 and J2215+5135 the most recently available ephemerides were optimised using \textsc{tempo2} to fit TOAs from the observations with epoch of periastron, T0, and projected semi-major axis of the orbit, A1, as free parameters, and the DMs were optimised using \textsc{pdmp} and used for incoherent dedispersion.\\
The results presented in P18 and P19 have demonstrated the value that can be gleaned from precise measurements of DM and scattering timescale of the pulses as a function of the pulsar's orbital phase. Consequently, the same model fitting method, described in Section 3.1 of P18, was utilised here to extract these quantities relative to the out-of-eclipse values. To summarise, this method consisted of generating 2-dimensional smooth model pulse templates (frequency channel vs. pulse phase) for each pulsar--telescope combination by stacking all available out-of-eclipse observations to give high signal-to-noise data, and smoothing these along both axes with a Savitzky-Golay filter \citep{sg64} to reduce noise fluctuations. For a given observation the corresponding smooth template was then fit to each sub-integration of data allowing the baseline level and pulse amplitude to vary freely, yielding a best-fit $\chi^2$ value from the residuals. The template was then dispersed and scattered with small increments in $\Delta$DM and $\Delta\tau$ and fit to the data again, yielding further $\chi^2$ values. This brute-force method essentially performed a grid search over pre-defined values of $\Delta$DM and $\Delta\tau$, taking the minimum $\chi^2$ fits to best represent the change in DM and $\tau$ of a sub-integration of data relative to the mean out-of-eclipse values. A key output from this method was the best-fit template amplitude for each sub-integration, as this is directly proportional to the detected flux density of the pulsar, thus allowing the relative flux density to be tracked throughout an eclipse. The resulting ``light-curves'' could then be used to measure the eclipse durations (see Section~\ref{sec: analysis_ecl_dur}).\\
Taking into account the possible variations in the mean pulse profiles over the long timescales covered by our observations, separate 2-dimensional template profiles were made for each widely spaced ($>1$\,week) set of observations. The step sizes used for the grid search over $\Delta$DM and $\Delta\tau$ are shown in Table~\ref{Table: dm_scat} for the new observations presented in this paper, decided primarily by the signal-to-noise of the observations.
\begin{table}
	\centering
	\caption[$\Delta$DM and $\Delta\tau$ step sizes used for template fits]{Step sizes in DM and $\tau$ used in the model fits to data observed at different centre frequencies. $\Delta\tau$ is expressed in units of the pulse period, $P$. $^{\text{a}}$ Fit performed over two simultaneously observed bands, 300--500\,MHz and 650--850\,MHz; $^{\text{b}}$ Fit performed over two simultaneously observed bands, 550--750\,MHz and 1260--1460\,MHz}
	\label{Table: dm_scat}
	\begin{tabular}{lcccc}
		\hline	
		PSR & Centre frequency & $\Delta$DM & $\Delta\tau$ & $P$ (ms) \\
		& (MHz) & (pc\,cm$^{-3}$) & & \\
		\hline
		J1816+4510 & 149 & $1\times10^{-4}$ & $0.005P$ & 3.19 \\
		J1816+4510 & 650 & $1\times10^{-3}$ & $0.005P$ & 3.19 \\
		B1957+20 & 149 & $1\times10^{-3}$ & $0.01P$ & 1.61 \\
		J1810+1744 & 575$^{\text{a}}$ & $1\times10^{-4}$ & $0.01P$ & 1.66 \\
		J2051$-$0827 & 149 & $2\times10^{-4}$ & $0.005P$ & 4.51 \\
		J2215+5135 & 1000$^{\text{b}}$ & $1\times10^{-3}$ & $0.01P$ & 2.61 \\
		\hline
	\end{tabular}
\end{table}\noindent

\subsection{Interferometric data}
The raw visibilities from the LOFAR observations were reduced in an averaging pipeline and the resulting outputs were stored in measurement sets with time intervals of 10\,s duration and frequency channels of $\sim49$\,kHz width. Further details of the LOFAR pipeline are available in \citet{h+10}. For these observations a short, $\sim10$\,min, scan of a dedicated flux calibrator source was performed immediately before and after the target observation. For B1957 these were 3C295 and 3C48, respectively, and for J1816, 3C295 was used for both pre- and post-target observations. The pre-processed data were calibrated using the LOFAR \textsc{prefactor} pipeline\footnote{https://github.com/lofar-astron/prefactor}, and images of the target field were made using \textsc{wsclean}\footnote{https://sourceforge.net/projects/wsclean/} \citep{omh+14} in time integrations of 2.5\,min and 5\,min for B1957 and J1816, respectively.\\
The earlier, 2014-Oct, observation of J1816 was instead planned to interleave calibrator observations throughout the long session, consisting of a repeating cycle of 30\,min on target and 7\,min on the flux calibrator 3C295. The pre-processing and calibration followed the same methods as used for PSR J1810+1744 in P18, similarly using \textsc{CASA}\footnote{https://casa.nrao.edu/} to image the pulsar in integrations of 5\,min. The repeating cycle of target--calibrator observations unfortunately led to small gaps in our data in both the ingress and egress of the image-plane eclipse.\\
From the resulting short-integration images, measurements of the detected pulsar flux density were made using \textsc{pybdsf}\footnote{http://www.astron.nl/citt/pybdsf/} to fit 2-dimensional Gaussians to the image pixel intensities at apparent source locations. For each observation a high signal-to-noise `detection image' was made by integrating all out-of-eclipse data, which was then used to initially identify the location of `islands' of emission used by \textsc{pybdsf} as indicators of source positions. These positions were then automatically used in \textsc{pybdsf} to fit model Gaussians in each short-integration image and extract the mean flux density, and $1\sigma$ uncertainty, over the integration time.

\subsection{Frequency dependent eclipse durations}\label{sec: analysis_ecl_dur}
Measuring the eclipse durations observed at different frequencies can give information on the dependence of the eclipses on the radiation frequency, and thus potentially give insight into the eclipse mechanism or properties of the eclipse medium. To achieve such measurements a method similar to those in P18 and \citet{bfb+16} was applied here. The eclipse ingress (egress) flux densities, $f$, as a function of orbital phase, $\phi$, were fit to a Fermi-Dirac type function, $f=\left[e^{\frac{\phi-p_1}{p_2}}+1\right]^{-1}$, with $p_1$ -- the orbital phase where the flux is half the out-of-eclipse value, and slope, $p_2$, as free parameters, via the method of least squares. To perform the least squares fit we made use of the \textsc{curve\_fit} function from \textsc{scipy}'s \textsc{optimize} package\footnote{https://www.scipy.org/}. Prior to the fits, the measured flux densities were normalised so that the mean out-of-eclipse flux density was unity. The normalisation was applied to ingress and egress data separately as often the out-of-eclipse means differed pre- and post-eclipse due the majority of the data having not been formally flux calibrated, meaning that slowly varying trends as a result of, for example, telescope pointing elevation, could still remain. This independent normalisation would also mask any differences in the average flux density intrinsic to the binary pre- and post-eclipse, however since these would be difficult to disentangle from the aforementioned instrumental effects, they were ignored. Notably, there were no systematic differences in the average out-of-eclipse flux densities either side of the eclipses. As the functional form used to model each eclipse edge was a one-sided `step-like' function, we were required to mask the opposite side of the eclipse while fitting an individual ingress or egress to avoid this biasing the residuals between the data and model in the least squares fits.\\
Using the fitted parameters, we defined the eclipse duration as the full-width at half of the out-of-eclipse flux density, i.e. $\Delta\phi_{\text{eclipse}} = p_{1,\text{eg}} - p_{1,\text{in}}$, where the subscripts `in' and `eg' refer to fits to eclipse ingress and egress, respectively. To avoid eclipse-to-eclipse variability from contaminating an individual eclipse duration measurement, we only calculated this metric for those observations that cover entire eclipses. Further, due to the observed asymmetry in many low-frequency eclipses we also measured durations for the eclipse ingresses and egresses separately; defined as $\Delta\phi_{\text{in/eg}} = \left|0.25 - p_{1,\text{in/eg}} \right|$, i.e. the orbital phase difference between ingress or egress and inferior conjunction of the companion, $\phi=0.25$.\\
Uncertainties in these metrics were taken to be the $1\sigma$ uncertainty in $p_1$ from the least squares fit for the ingress and egress durations, and the quadrature sum of the $1\sigma$ uncertainties in $p_{1,\text{in}}$ and $p_{1,\text{eg}}$ for the full eclipse duration. Unless otherwise stated -- see Section~\ref{sec: J2215} for PSR J2215+5135 -- these formal uncertainties were assumed throughout further analyses, meaning that each individual measurement made no attempt to account for temporal variability in the eclipses.\\
As in previous analyses \citep[P18;][]{ntt+90,fbb+90,rt91,bfb+16} we attempted to glean insight from these results by fitting power law curves to the measured durations of the form $\Delta\phi = A (\nu/150)^{-\alpha}$, where $\nu$ is the observing frequency and $A$ and $\alpha$ are free parameters to be fit, representing the effective location and steepness of the curve, respectively. Similarly to the Fermi-Dirac fits explained above, we utilised the \textsc{curve\_fit} function to perform least squares fits. Separate power laws were fit in the case that clear temporal variability between eclipses was present -- see Section~\ref{sec: J1816} for J1816. It is the resulting $\alpha$ parameters that we consider here -- the $A$ parameter accounts only for systematic variation of the eclipse width at 150\,MHz between different pulsar systems -- and these are presented in Table~\ref{Table: psr_params}. In Tables~\ref{Table: J1816_ecl} and \ref{Table: pwrlaws} the measured eclipse durations, fitted power law coefficients and corresponding uncertainties are given, with the intention for these to be utilised in future studies.

\section{Eclipse characterisation}\label{sec: ecl_obs}
\subsection{PSR J1816+4510}\label{sec: ecl_obs_J1816}
\begin{figure*}
	\includegraphics[width=.85\textwidth]{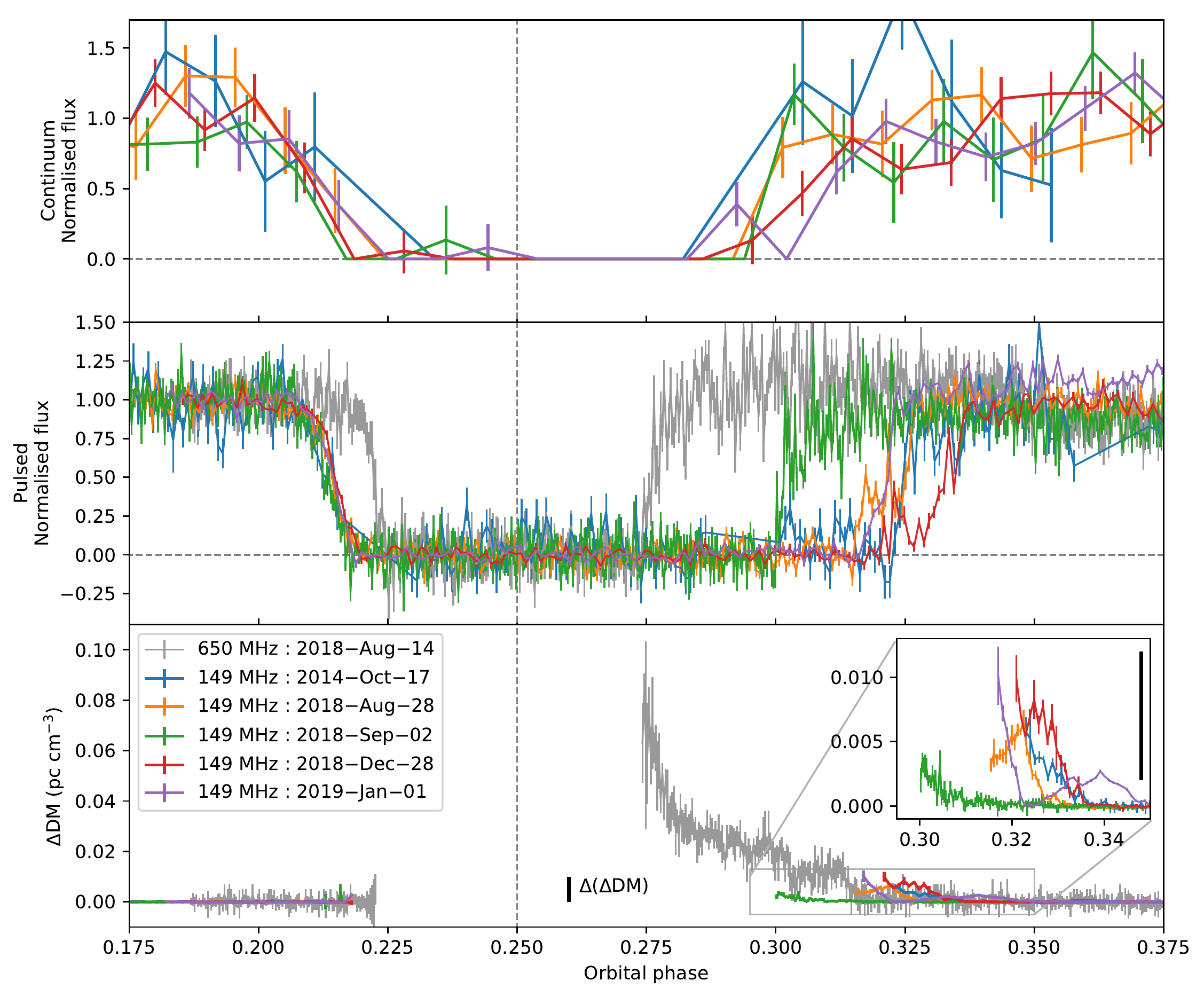}
	\caption[DM and flux density variations for all observed eclipses of PSR J1816+4510]{Measurements of the radio emission of PSR J1816+4510 throughout the eclipse region from simultaneous beamformed and imaging observations at 149\,MHz. \textit{Top}: Un-pulsed flux densities from continuum images, with each normalised so that the out-of-eclipse mean flux density is unity. The horizontal dashed line corresponds to the detection limit of the telescope. \textit{Middle}: Pulsed flux densities from beamformed observations, again normalised so that the out-of-eclipse mean flux density is unity. \textit{Bottom}: Deviation from mean out-of-eclipse dispersion measures for the same set of observations. The grey curves in the middle and bottom panels are measured from beamformed data observed at 650\,MHz. The black bar in the bottom panel, marked $\Delta(\Delta$DM), represents the minimum change in $\Delta$DM within a single sub-integration time that would be sufficient to smear out pulsations (see Section~\ref{sec:mechanisms}); the same line is shown in the inset plot. Colours are consistent between panels, and error bars represent $1\sigma$ uncertainties from the simultaneous DM and scattering fits, as explained in Section \ref{sec: analysis_bf}.}
	\label{fig:J1816_flux_dm}
\end{figure*}

\begin{figure*}
	\includegraphics[width=0.85\textwidth]{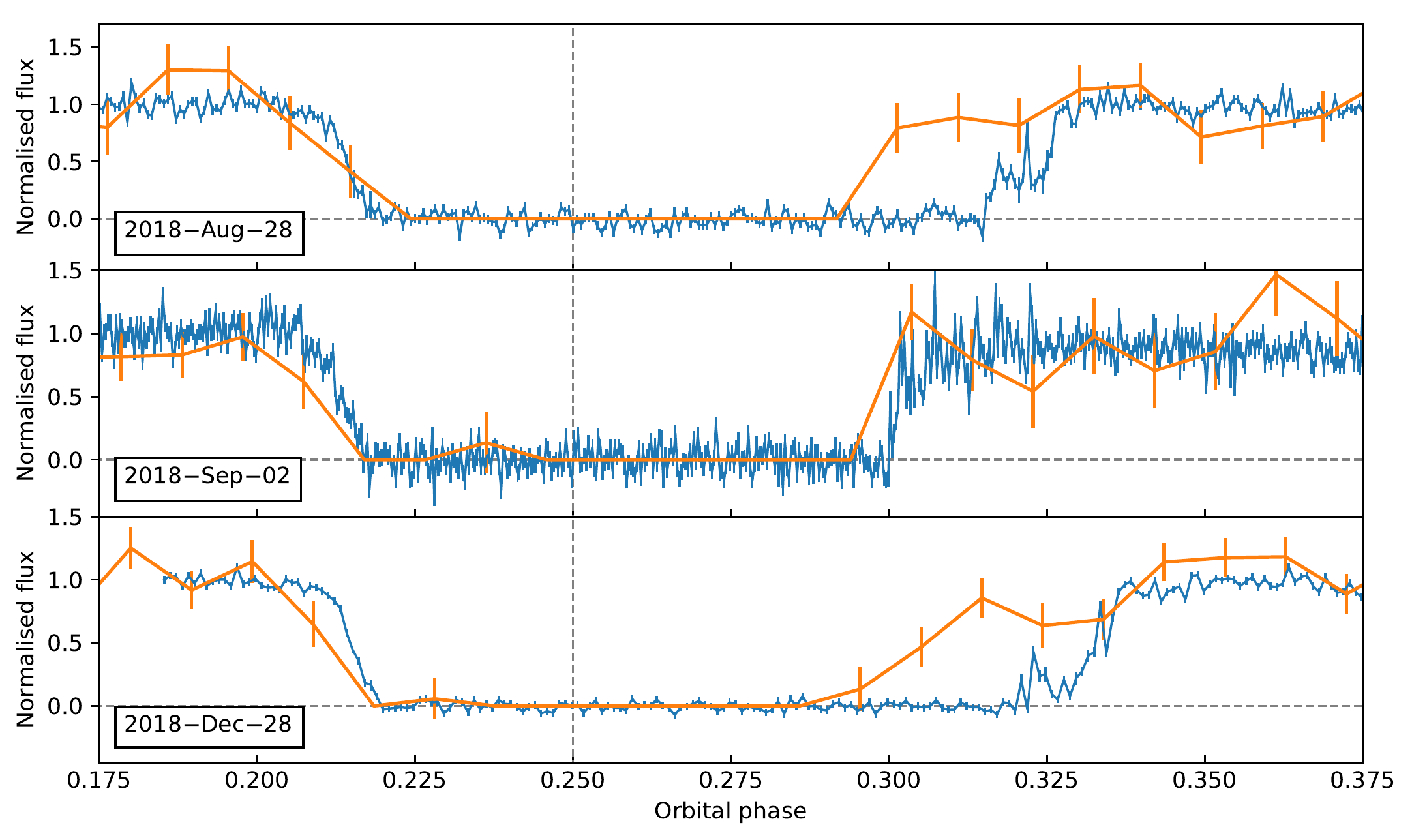}
	\caption[Over-plotted pulsed and un-pulsed eclipses of PSR J1816+4510]{Un-pulsed flux densities of PSR J1816+4510 from continuum images (orange), and corresponding pulsed flux densities (blue) for three separate eclipses. Flux densities are normalised so that the out-of-eclipse mean flux density is unity. The horizontal dashed line corresponds to the detection limit of the telescope.}
	\label{fig:J1816_img_vs_bf}
\end{figure*}

For J1816, Fig.~\ref{fig:J1816_flux_dm} shows the collective measurements of the \textit{un-pulsed} flux density obtained from images, each integrated over 5\,mins, and the \textit{pulsed} flux density and deviation in DM, output from the template fits to the beamformed data. Both the pulsed and un-pulsed flux densities have been normalised such that the mean out-of-eclipse flux densities are unity, allowing a clear comparison of the eclipse effects.\\
\noindent
At first sight some of the apparently now common features of low-frequency spider eclipses are visible. The eclipse is centred to a later orbital phase than inferior conjunction of the companion -- phase 0.25 -- with the 149\,MHz eclipse lasting up to 3 times longer after phase 0.25 than before it, while the higher-frequency eclipse is much more symmetric. There is a relatively stable ingress with little excess DM and a much more erratic egress with large DM fluctuations. Such features are generally attributed to a `tail' of material that is swept behind the companion by Coriolis forces as it moves around the orbit \citep[e.g.][]{fbb+90}, and it appears reasonable to infer that the same phenomenon is responsible here. Equally clear is the complete attenuation -- at least by a factor of 5 based on the signal-to-noise of the out-of-eclipse detections -- of the continuum, un-pulsed flux in the eclipse region. As far as we are aware this has been observed in all published low-frequency imaging observations of spider eclipses thus far, although this only amounts to four other systems \citep[P18;][]{fg92,rrb+15,bfb+16}. Taking the definition of an eclipse to be a persistent region of measured flux density less than or equal to the average flux density away from the nominal eclipse phases, we find that the duration of the eclipses show significant frequency dependence (see also Section~\ref{sec: J1816}). The variable low-frequency eclipses last for $\sim9$--12\% of the orbit, while the higher-frequency eclipse covers only $\sim5$\%. Notably, if the, as yet unconstrained, inclination of the orbit is near $90^{\circ}$ then the Roche lobe would span $\sim7$\% of the orbit, assuming the parameter values in Table~\ref{Table: psr_params}, and as such the 650\,MHz flux would penetrate the star if it were to fill its Roche lobe in this case. This suggests that either the orbit is not edge-on or the companion does not fill its Roche lobe. For a Roche lobe filling companion, the inclination of the orbit would need to be $\lesssim81^{\circ}$ for the line-of-sight `slice' across the Roche lobe to be $\lesssim5$\% of the orbit.\\
Under closer inspection a particularly interesting feature emerges: comparing the continuum eclipses to the pulsed eclipses it can be seen that the orbital phases of the ingresses are regularly consistent with one another -- also seen in PSR J1810+1744 (P18) -- however in all but one eclipse egress the pulsed eclipse extends to a significantly later phase than the corresponding continuum eclipse. This is more readily shown in Fig.~\ref{fig:J1816_img_vs_bf}, where the pulsed and corresponding un-pulsed flux density measurements are over-plotted for three of the eclipses. This provides clear evidence for the presence of an eclipse mechanism that removes pulsar flux from our line of sight, then upon approaching egress transitions to one that merely smears out pulsations while not removing flux from the line of sight. This is the first ever such observational evidence for any spider pulsar, and is discussed further in Section~\ref{sec:mechanisms}.\\
\noindent
The inset plot of Fig.~\ref{fig:J1816_flux_dm} magnifies the egress DM deviations, which show that the excess material in the tail is not smooth on these scales, and often slowly undulating structures are present that suggest relatively large scale clumps. Interestingly, in simplified two-dimensional hydrodynamical models of the outflow in B1957, \citet{tb91} show that time-dependent, large scale, density enhancements can indeed form in the tail of the outflow. The timescale of large variations in the tail material is constrained by our observations to be $<4$\,days ($\sim10$ orbits); notably the early-exiting egress, with relatively little DM fluctuations, on 2018-Sep-02 shows no similarities to that observed just 5\,days previously. Comparing the eclipses observed in 2018 to that in 2014, we see no significant difference that could suggest long-term variability, although the extremely sparse sampling of eclipses over this time-span means that it cannot be ruled out.\\
At 650\,MHz the changes in DM at eclipse boundaries are still highly asymmetric and no significant increase is seen in the ingress prior to loss of flux. This appears to be consistent with the TOAs presented in Fig.~10 of \citet{slr+14}, where a sharp rise in ingress DM is only detected at 820\,MHz and 1500\,MHz, probing further into the eclipse. In addition, the general shape of the $\Delta$DM pattern, which is relatively smooth on these magnitudes, is very well replicated by those seen in B1957 at a similar frequency \citep{rt91}.

\subsection{PSR B1957+20}\label{sec: ecl_obs_B1957}
\begin{figure*}
	\includegraphics[width=.85\textwidth]{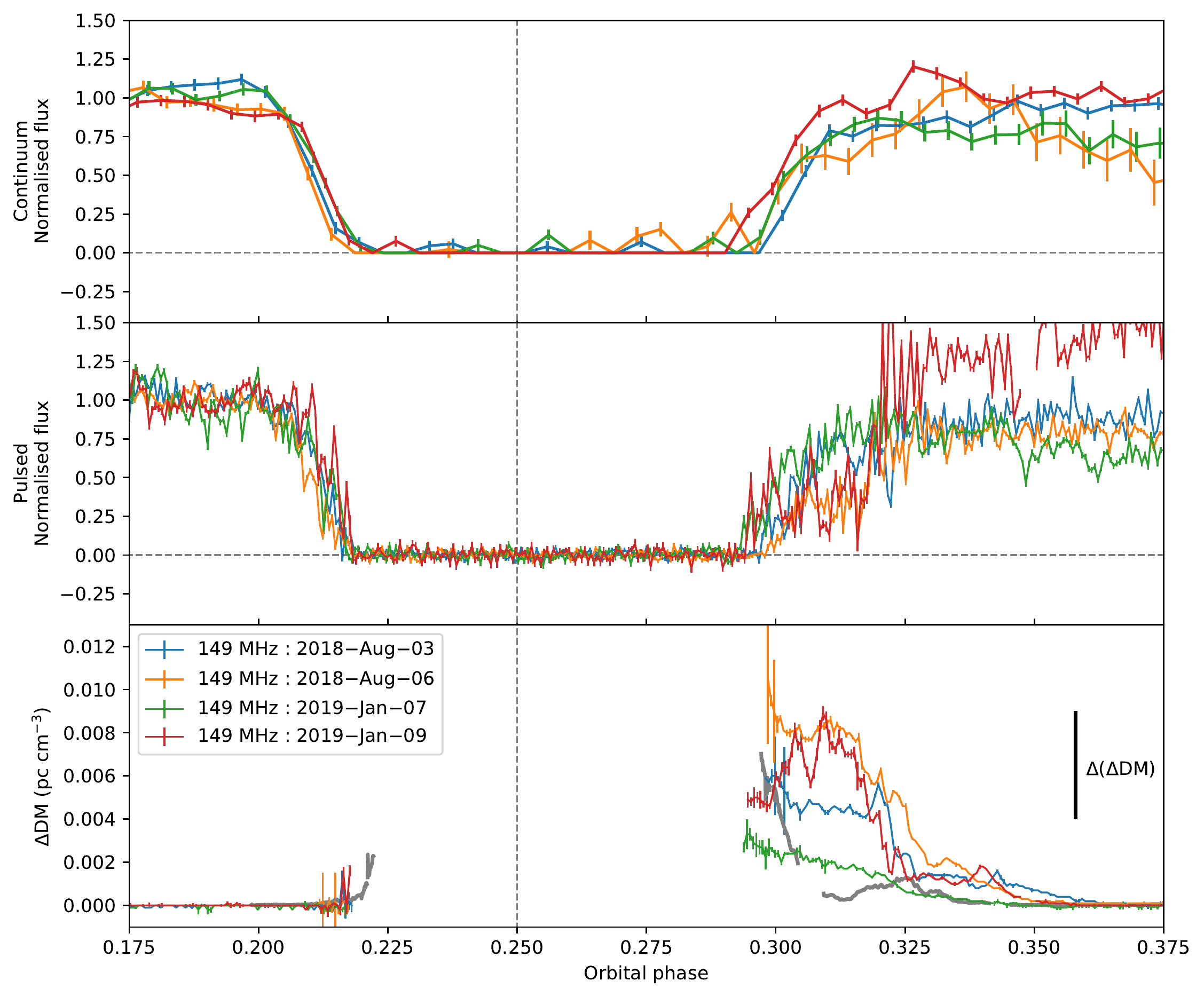}
	\caption[DM and flux density variations for all observed eclipses of PSR B1957+20]{Measurements of the radio emission of PSR B1957+20 throughout the eclipse region from simultaneous beamformed and imaging observations at 149\,MHz. \textit{Top}: Un-pulsed flux densities from continuum images, with each normalised so that the out-of-eclipse mean flux density is unity. The horizontal dashed line corresponds to the detection limit of the telescope. \textit{Middle}: Pulsed flux densities from beamformed observations, again normalised to unity. \textit{Bottom}: Deviation from mean out-of-eclipse dispersion measures for the same set of observations. Overplotted in grey are the $\Delta$DMs observed at 335\,MHz as published in \citet{myc+18}. The black bar in the bottom panel, marked $\Delta(\Delta$DM), represents the minimum change in $\Delta$DM within a single sub-integration time that would be sufficient to smear out pulsations (see Section~\ref{sec:mechanisms}). Colours are consistent between panels, and error bars represent $1\sigma$ uncertainties from the simultaneous DM and scattering fits, as explained in Section \ref{sec: analysis_bf}.}
	\label{fig:B1957_flux_dm}
\end{figure*}

Fig.~\ref{fig:B1957_flux_dm} presents the same measurements of un-pulsed flux density, pulsed flux density and deviation of DM, now for B1957 as a function of its orbit. The results are shown for four 149\,MHz eclipses, with intervals between observations ranging from a few days to 5\,months. Given the higher signal-to-noise of B1957 observations relative to those of J1816, the integration time of each image was reduced to 2.5\,mins, allowing $\sim5$ flux density samples in each eclipse ingress and egress. This increased resolution permits stringent tests on the orbital phases of pulsed versus un-pulsed eclipses, and shows that the two are closely matched in both the times of ingress and of egress, suggesting that the low-frequency eclipse occurs solely as a result of removal of flux from the line of sight. On the other hand, in the 2018-Aug-06 and 2019-Jan-09 egresses -- where $\Delta$DM is largest -- there is some reduction in pulsed relative to un-pulsed flux density part way through the egress. This is likely explained by low-level scattering causing broadening of the pulsations. Indeed, our fits of $\Delta$DM and $\Delta\tau$ are consistent with this being the case, however the full pulsed flux density is not accounted for here as the pulse profile at 149\,MHz has a duty cycle of $\sim100$\%, thus any scatter-broadening of the profile effectively smears the flux into the arbitrary baseline level, which is automatically subtracted in pre-processing of the data. It could well be the case that a more severe effect of the same mechanism is responsible for the extended egresses in J1816, as this pulsar too has a $\sim100$\% duty cycle.\\
\noindent
Taking into account the typical signal-to-noise of the out-of-eclipse images, the flux density during eclipse is reduced by at least a factor of $\sim25$ and is consistent with zero flux. This is in agreement with the observation at 330\,MHz in \citet{fg92}, whereby a single image integrated over the entire eclipse region constrained the flux density of B1957 to $(1\pm3)$\,mJy, whereas this value was found to be $(38\pm3)$\,mJy when integrated over the full observation.\\
The patterns in $\Delta$DM that we measure are consistent with the measurements at $>300$\,MHz made in the few years following the initial discovery \citep{fbb+90,rt91}, and the corresponding interpretations of a swept back, low-density tail of material trailing behind the companion still appear to be applicable. Similar to J1816, and at higher frequencies in B1957, there are slowly undulating patterns in the DM, however our high signal-to-noise measurements also reveal rapid fluctuations of the order of $10^{-3}$\,pc\,cm$^{-3}$ on 20\,s timescales. This contrasts the $\sim10^{-5}$\,pc\,cm$^{-3}$ fluctuations on 2\,s timescales reported in \citet{myc+18}, for which the corresponding eclipse, observed at 325\,MHz in 2014-Jun, is over-plotted on our observations in grey in the bottom panel of Fig.~\ref{fig:B1957_flux_dm}.

\subsubsection{Out-of-eclipse variability}\label{sec: B1957_toas}
\begin{figure*}
	\includegraphics[width=\textwidth]{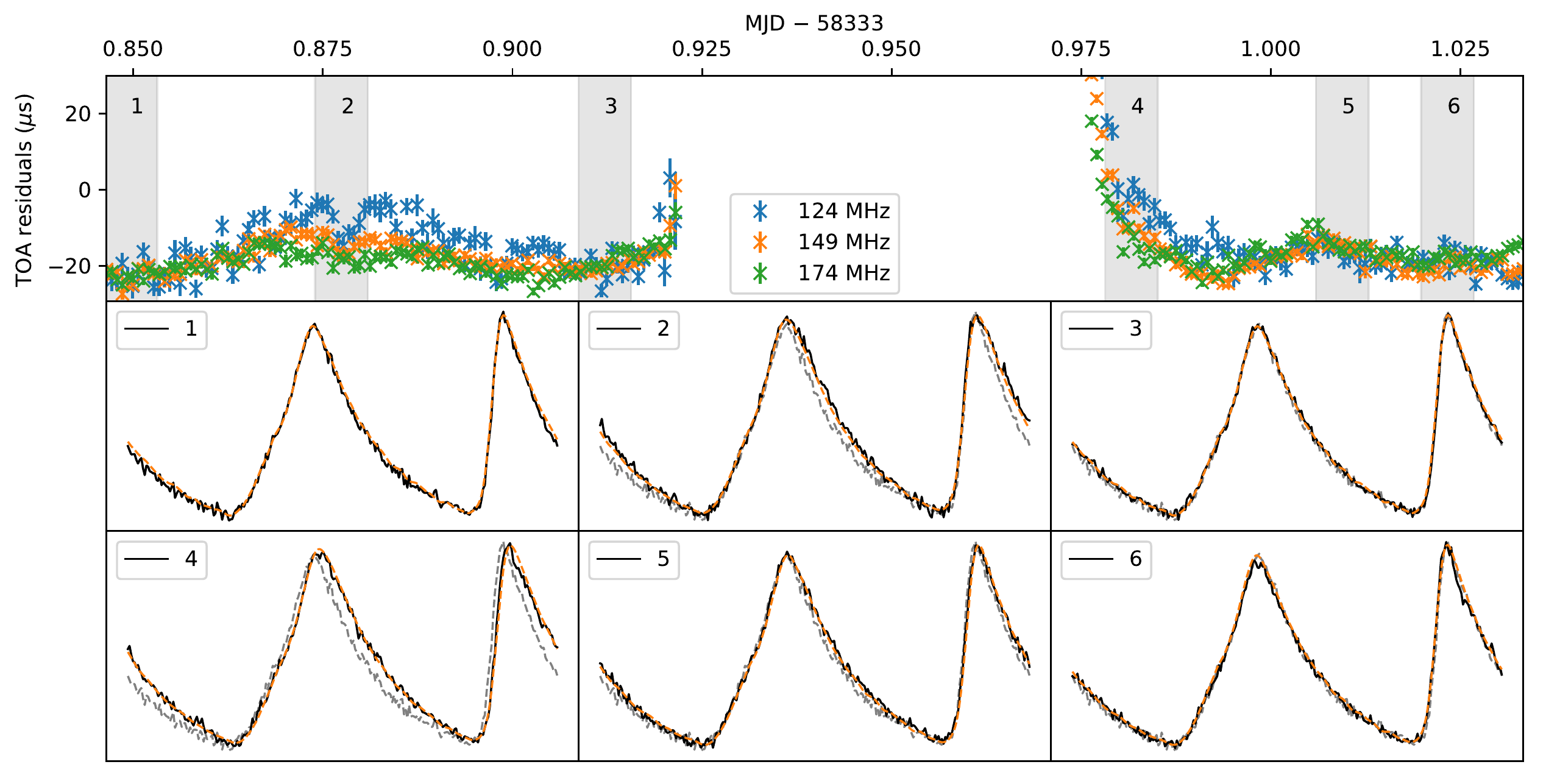}
	\caption[Pulse profile broadening and timing structure in PSR B1957+20 -- 2018-Aug-03]{\textit{Top}: Measured time-of-arrival of pulses in 3 non-overlapping sub-bands relative to the timing model for the observation of PSR B1957+20 on 2018-Aug-03. Data from the eclipse region has been removed for clarity. \textit{Middle/Bottom}: Pulse profiles integrated over the full frequency band and 10\,mins (black) at the indicated times. The grey, dashed profiles represent the baseline profile (profile 1), and the orange profiles show the smooth pulse template used to model variations in DM and $\tau$ after the best-fit values of phase shift, $\Delta$DM and $\Delta\tau$ have been applied.}
	\label{fig:toa_profs1}
\end{figure*}

\begin{figure*}
	\includegraphics[width=\textwidth]{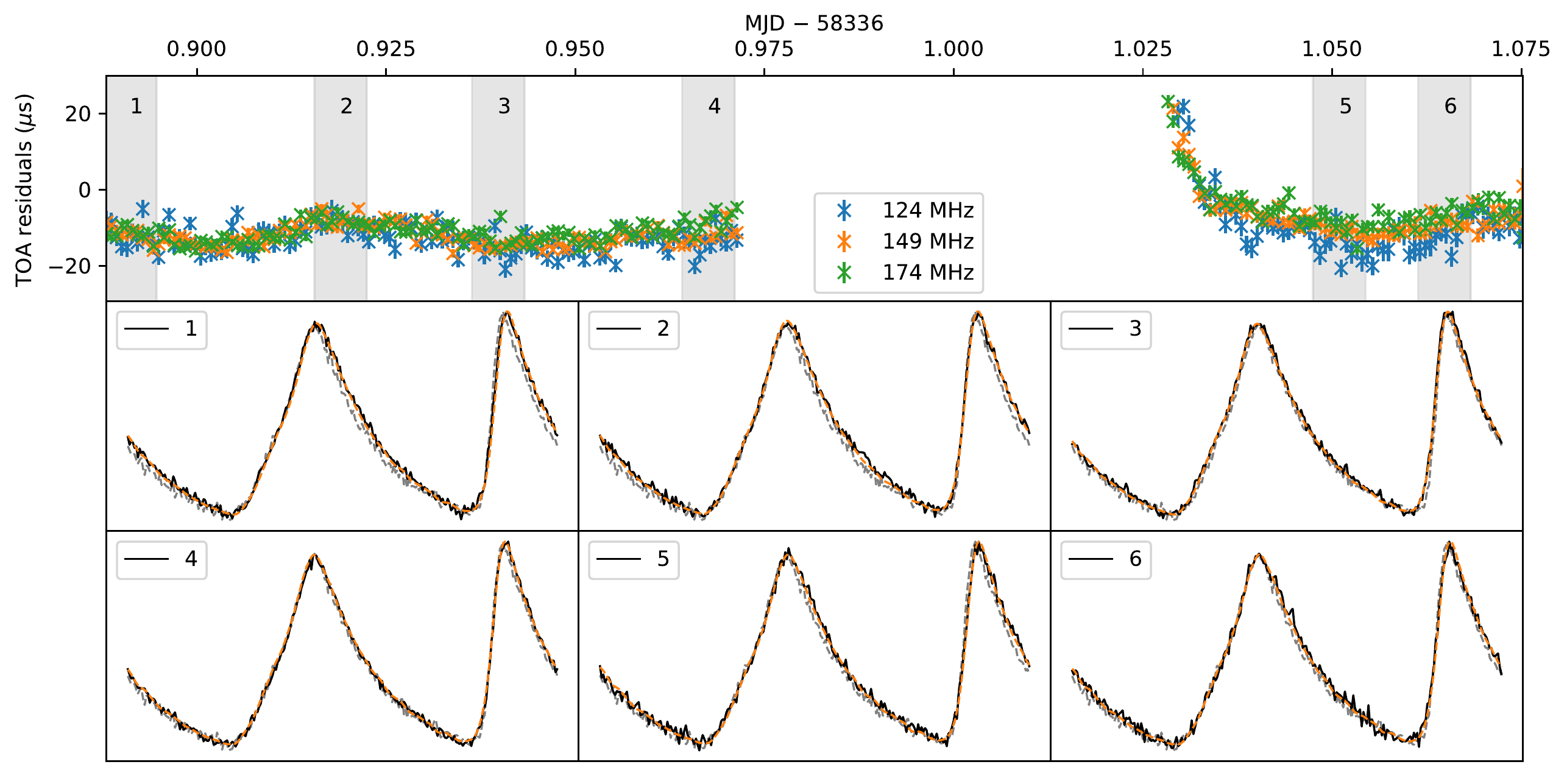}
	\caption[Pulse profile broadening and timing structure in PSR B1957+20 -- 2018-Aug-06]{\textit{Top}: Measured time-of-arrival of pulses in 3 non-overlapping sub-bands relative to the timing model for the observation of PSR B1957+20 on 2018-Aug-06. Data from the eclipse region has been removed for clarity. \textit{Middle/Bottom}: Pulse profiles integrated over the full LOFAR frequency band and 10\,mins (black) at the indicated times. The grey, dashed profiles represent the baseline profile (profile 1 from Fig.~\ref{fig:toa_profs1}), and the orange profiles show the template after the best-fit values of phase shift, $\Delta$DM and $\Delta\tau$ have been applied.}
	\label{fig:toa_profs2}
\end{figure*}

\begin{figure*}
	\includegraphics[width=\textwidth]{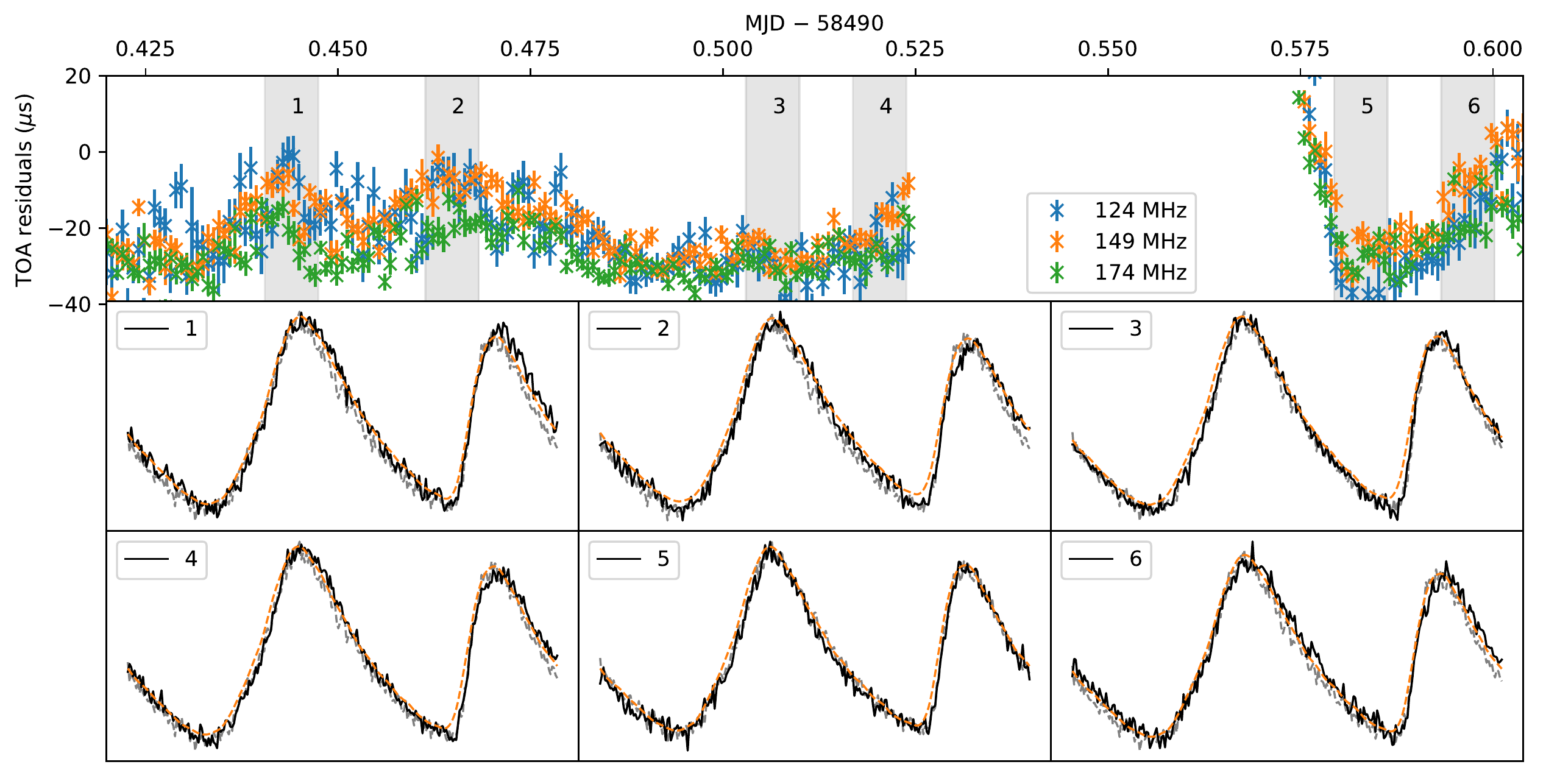}
	\caption[Pulse profile broadening and timing structure in PSR B1957+20 -- 2019-Jan-07]{\textit{Top}: Measured time-of-arrival of pulses in 3 non-overlapping sub-bands relative to the timing model for the observation of PSR B1957+20 on 2019-Jan-07. Data from the eclipse region has been removed for clarity. \textit{Middle/Bottom}: Pulse profiles integrated over the full frequency band and 10\,mins (black) at the indicated times. The grey, dashed profiles represent the baseline profile (profile 1 from Fig.~\ref{fig:toa_profs4}), and the orange profiles show the template after the best-fit values of phase shift, $\Delta$DM and $\Delta\tau$ have been applied.}
	\label{fig:toa_profs3}
\end{figure*}

\begin{figure*}
	\includegraphics[width=\textwidth]{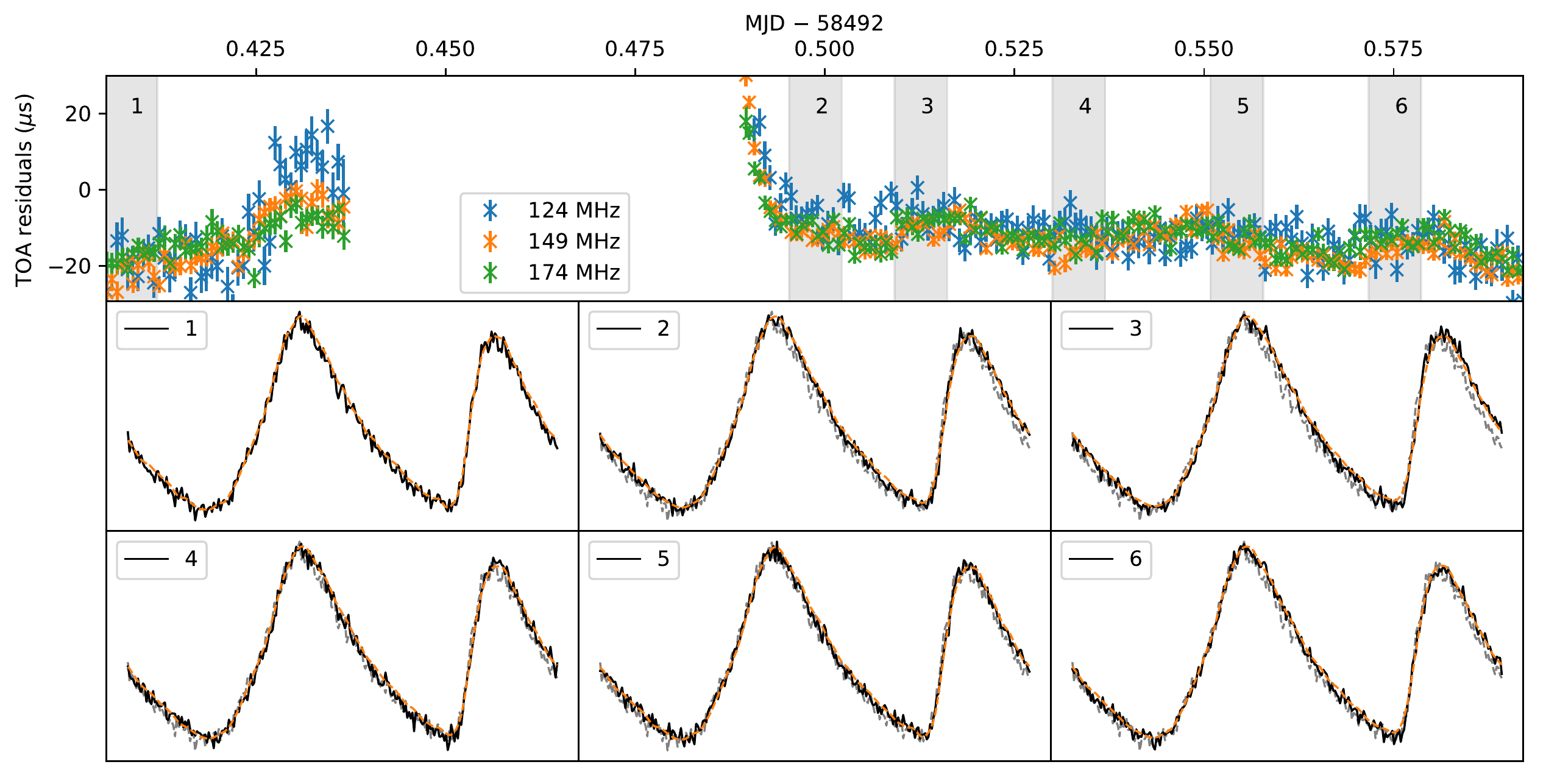}
	\caption[Pulse profile broadening and timing structure in PSR B1957+20 -- 2019-Jan-09]{\textit{Top}: Measured time-of-arrival of pulses in 3 non-overlapping sub-bands relative to the timing model for the observation of PSR B1957+20 on 2019-Jan-09. Data from the eclipse region has been removed for clarity. \textit{Middle/Bottom}: Pulse profiles integrated over the full frequency band and 10\,mins (black) at the indicated times. The grey, dashed profiles represent the baseline profile (profile 1), and the orange profiles show the template after the best-fit values of phase shift, $\Delta$DM and $\Delta\tau$ have been applied.}
	\label{fig:toa_profs4}
\end{figure*}

As detailed in Section~\ref{sec: analysis_bf}, we measured pulse TOAs to fit to a timing model of B1957 in order to refine the orbital parameters. Upon inspection of the TOA residuals -- the discrepancies between the expected TOAs from the model, and the measured TOAs from the data -- there was unusual structure with magnitudes of $\sim20$\,$\mu$s, and timescales of minutes--hours that could not be accounted for with the usual pulsar rotation and orbital parameters. To investigate the possibility of a frequency-dependent mechanism being responsible for the variations -- e.g. scattering or DM -- we split the LOFAR band into 3 sub-bands and obtained TOAs for each. The TOA residuals relative to the timing model are shown in Figs.~\ref{fig:toa_profs1},~\ref{fig:toa_profs2},~\ref{fig:toa_profs3} and~\ref{fig:toa_profs4} for the four observations. The TOA measurements alone are not conclusive, but do appear to show frequency structure in some of the larger delays. Taking this further, the DM and $\tau$ template fitting method (Section~\ref{sec: analysis_bf}) was applied with increments of $\Delta$DM and $\Delta\tau$ small enough to allow multiple grid search elements across the observed $\sim20$\,$\mu$s delays. Two templates were generated, one for the 2018-Aug observations, and one for the 2019-Jan observations, integrating over $\sim30$\,mins of data to gain signal-to-noise while avoiding integrating over too much of the delays that would act to broaden the template profiles. In addition to the standard template fitting method, a third grid search dimension was included to allow positive or negative phase shifts of the template, simultaneously in all frequency channels, in order to model frequency-independent TOA structure.\\
\noindent
\noindent
\noindent
\noindent
The results of the fits suggest unrealistic interchanges between phase shifts, $\Delta$DM and $\Delta\tau$ from one integration to the next. This is likely a consequence of the degeneracy between these effects for small perturbations in the presence of noise, and will also be influenced by the difficulty in making a `baseline' template profile for the fits when the profile variations are continuous throughout the orbit, thus meaning that at least some subset of the variations must be integrated over in the creation of the template. However, as shown in Figs.~\ref{fig:toa_profs1},~\ref{fig:toa_profs2},~\ref{fig:toa_profs3} and~\ref{fig:toa_profs4}, it is evident that the pulse profiles -- integrated over frequency -- become broadened on the latter edges in regions of relatively delayed TOA measurements, strongly suggesting that either increased scattering, DM, or some combination of the two does indeed occur.\\
It is difficult to isolate the location along the line of sight of the medium responsible for these suggested propagation variations, however should it be related to the outflow from the companion then it would be the first such indication of excess material located so far from the primary eclipse in this well-studied system. Indeed, it would not be unexpected that such structure has not been identified previously as the effects of scattering and DM scale with frequency as $\nu^{-4}$ and $\nu^{-2}$, respectively, meaning that they would be difficult to detect in the previously published observations of B1957, typically at frequencies $>300$\,MHz.

\section{Eclipse frequency dependence}\label{sec: freq_dependence}
Here we investigate the dependence of the eclipse durations on observing wavelength, considering also a number of other spider pulsars for which data is available.

\subsection{PSR J1816+4510}\label{sec: J1816}
\begin{figure}
	\centering
	\includegraphics[width=0.5\columnwidth]{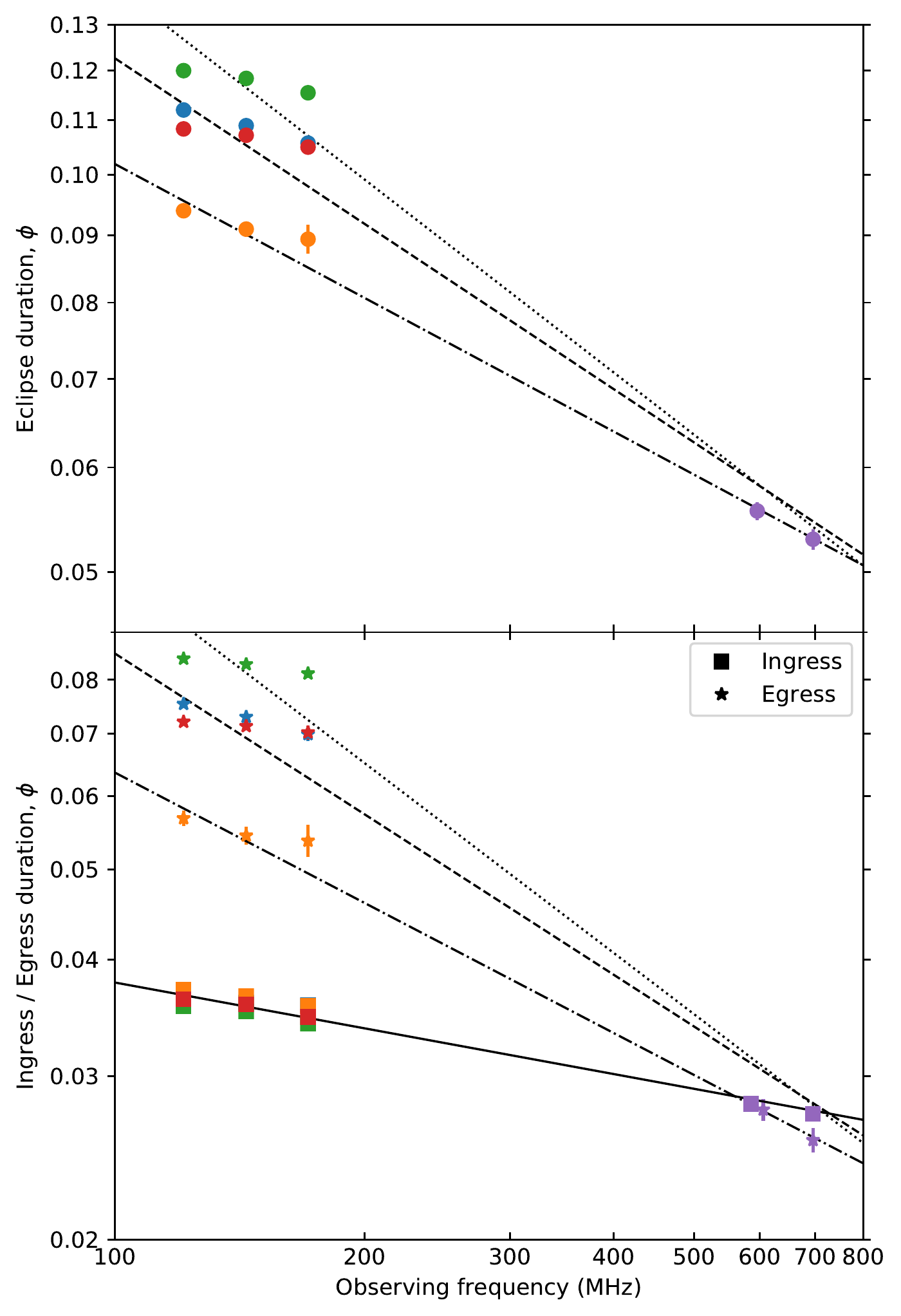}
	\caption[Eclipse durations versus frequency for PSR J1816+4510]{Eclipse durations (orbital phase range in which detected flux density is less than half that of the average out-of-eclipse level) versus frequency for PSR J1816+4510. Colours are consistent between panels and represent separate eclipses. \textit{Top}: Full eclipse durations. Fitted power law lines have slopes $\alpha_{\text{full}}=0.34$ (dash-dot) -- fit to orange LOFAR and purple GMRT points; $\alpha_{\text{full}}=0.42$ (dashed) -- fit to red and blue LOFAR and purple GMRT points; $\alpha_{\text{full}}=0.49$ (dotted) -- fit to green LOFAR and purple GMRT points. \textit{Bottom}: Ingress and egress durations. Fitted power law lines have slopes $\alpha_{\text{in}}=0.16$ (solid) -- fit to all ingress points; $\alpha_{\text{eg}}=0.46$ (dash-dot) -- fit to orange LOFAR and purple GMRT points; $\alpha_{\text{eg}}=0.58$ (dashed) -- fit to red and blue LOFAR and purple GMRT points; $\alpha_{\text{eg}}=0.68$ (dotted) -- fit to green LOFAR and purple GMRT points. The low-frequency egress points appear to follow a different power law to those at high-frequency (see Section \ref{sec: J1816}).}
	\label{fig:J1816_pwrlaw}
\end{figure}

The measured eclipse durations and corresponding best-fit power laws for J1816 are shown in the top panel of Fig.~\ref{fig:J1816_pwrlaw}. With different coloured points representing separate eclipses, the strong influence of temporal variability of the eclipse properties on their apparent frequency dependence is immediately obvious for the measurements near 150\,MHz. Shown in the bottom panel are the separate ingress and egress durations, relative to inferior conjunction of the companion. These measured parameters are given in Table~\ref{Table: J1816_ecl}, along with the fitted power law parameters in Table~\ref{Table: pwrlaws}. The large variability in the egress durations at low-frequency was highlighted in Section~\ref{sec: ecl_obs_J1816}, with the analysis suggesting that different mechanisms were responsible for the latter part of the eclipse in comparison to the ingress and main duration, particularly for the extended egresses. As discussed further in Section~\ref{sec:mechanisms}, it appears to be most likely that the pulsed emission becomes smeared in time after propagating through a tenuous tail of material, causing the eclipses to be extended, which would have much less influence on higher-frequency emission for which typical pulse smearing effects (e.g. scattering or dispersion) would be less prominent. This picture is supported by the observed stability in the eclipse ingress, which would be unaffected by a trailing tail of material.\\
\noindent
\begin{table*}
	\centering
	\caption{Measured eclipse durations: full eclipse, $\Delta\phi_{\text{eclipse}}$, ingress, $\Delta\phi_{\text{in}}$, and egress, $\Delta\phi_{\text{eg}}$, along with formal $1\sigma$ uncertainties. Here the first 5 lines are shown; the full table can be found in the supplementary online material.}
	\label{Table: J1816_ecl}
	\begin{tabular}{lccccc}
		\hline
		PSR & Telescope & $\nu$\,(MHz) & $\Delta\phi_{\text{eclipse}}$ & $\Delta\phi_{\text{in}}$ & $\Delta\phi_{\text{eg}}$\\
		\hline
		J1816+4510 & LOFAR & 121 & $0.1120\pm0.0008$ & $0.0366\pm0.0004$ & $0.0754\pm0.0006$\\
		& & & $0.0939\pm0.0012$ & $0.0371\pm0.0005$ & $0.0568\pm0.0011$\\
		& & & $0.1200\pm0.0008$ & $0.0357\pm0.0004$ & $0.0843\pm0.0007$\\
		& & & $0.1084\pm0.0006$ & $0.0363\pm0.0003$ & $0.0721\pm0.0005$\\
		& & 144 & $0.1090\pm0.0008$ & $0.0360\pm0.0003$ & $0.0730\pm0.0007$\\
		\hline
	\end{tabular}
\end{table*}\noindent
\begin{table*}
	\centering
	\caption{Best-fit power law parameters for eclipse durations, when fit to full eclipse durations, ingress durations and egress durations as a function of frequency, $\nu$, i.e. $\Delta\phi = A (\nu / 150)^{-\alpha}$. $^{\text{a}}$~P18. Here the first 5 lines are shown; the full table can be found in the supplementary online material.}
	\label{Table: pwrlaws}
	\begin{tabular}{lcccccc}
		\hline
		PSR & \multicolumn{2}{c}{Full eclipse} & \multicolumn{2}{c}{Ingress} & \multicolumn{2}{c}{Egress} \\
		& $A$ & $\alpha$ & $A$ & $\alpha$ & $A$ & $\alpha$ \\
		\hline
		J1810+1744 & $0.129\pm0.001$ & $0.22\pm0.02$ & $0.050\pm0.001$ & $0.09\pm0.01$ & $0.076\pm0.001$ & $0.29\pm0.02$\\
		$^{\text{a}}$ & $0.127\pm0.001$ & $0.41\pm0.02$ & $0.049\pm0.001$ & $0.41\pm0.05$ & $0.075\pm0.001$ & $0.35\pm0.03$\\
		J1816+4510 & $0.114\pm0.003$ & $0.49\pm0.06$ & $0.035\pm0.001$ & $0.16\pm0.01$ & $0.078\pm0.003$ & $0.68\pm0.11$\\
		& $0.103\pm0.002$ & $0.42\pm0.04$ & & & $0.067\pm0.002$ & $0.58\pm0.07$\\
		& $0.089\pm0.001$ & $0.34\pm0.02$ & & & $0.053\pm0.001$ & $0.46\pm0.02$\\
		\hline
	\end{tabular}
\end{table*}\noindent
The ingress power law gives a good fit across all four 149\,MHz eclipses and the single 650\,MHz eclipse, suggesting that any effect of time variability is negligible here. However, for the more complex egress we take the most shallow power law to give the most reliable indicator of the main bulk of the eclipse medium, for which the pulsed and unpulsed eclipses were consistent in duration, suggesting only a small amount of eclipse extension due to pulse smearing. Indeed, this gives a much better fit across both the LOFAR and GMRT bandwidths (reduced $\chi^2 = 2$) relative to the extended eclipses, for which a single power law is evidently not sufficient (reduced $\chi^2 = 34$ and 80 for the dashed and dotted power laws, respectively).\\
As a consequence of the inconsistency of the egress frequency dependence across the low-frequencies with that of the high-frequencies, we also performed fits to the data $<200$\,MHz only. As for the full frequency coverage data, we fitted 3 separate power laws, using the two eclipses in blue and red in a single fit. The resulting power law exponents are presented in Table~\ref{Table: J1816_pwrlaw}, along with their weighted mean, and are discussed further in the context of B1957 below.
\begin{table}
	\centering
	\caption[Eclipse duration power law fits for PSR J1816+4510]{Best-fit power law exponents for eclipse durations in PSR J1816+4510, when fit to \textit{LOFAR data only}, i.e. $\Delta\phi_{\text{eclipse}} \propto \nu^{-\alpha}$, for $100 < \nu < 200$\,MHz. The weighted means of $\alpha_{\text{full}}$ and $\alpha_{\text{eg}}$ are given in the bottom row.}
	\label{Table: J1816_pwrlaw}
	\begin{tabular}{cc}
		\hline
		$\alpha_{\text{full}}$ & $\alpha_{\text{eg}}$ \\
		\hline
		$0.11\pm0.02$ & $0.10\pm0.02$ \\
		$0.11\pm0.05$ & $0.13\pm0.08$ \\
		$0.16\pm0.03$ & $0.19\pm0.06$ \\
		\hline
		$0.12\pm0.02$ & $0.11\pm0.02$ \\
		\hline
	\end{tabular}
\end{table}\noindent

\subsection{PSR B1957+20}\label{sec: B1957}
\begin{figure}
	\centering
	\includegraphics[width=0.5\columnwidth]{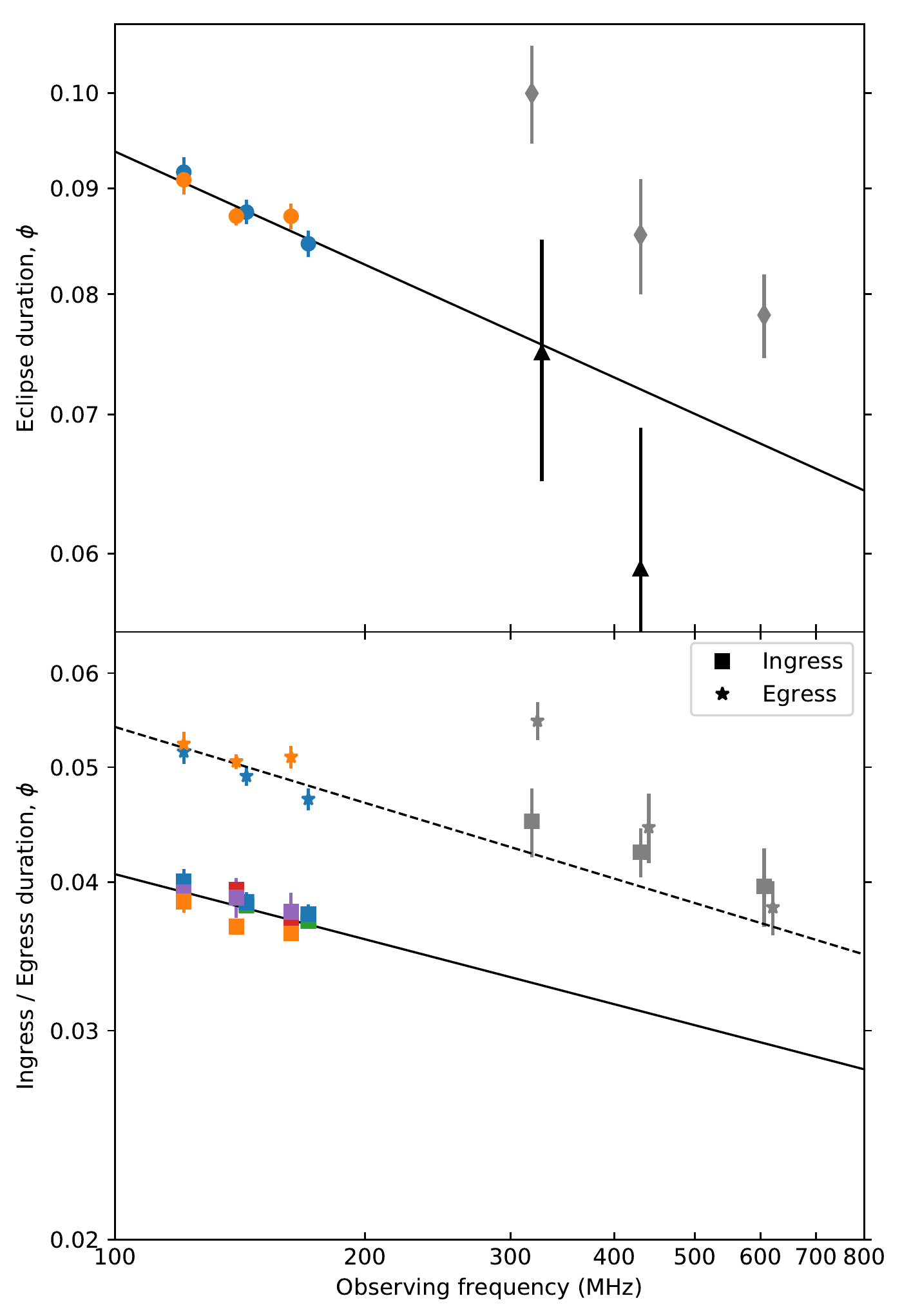}
	\caption[Eclipse durations versus frequency for PSR B1957+20]{Eclipse durations versus frequency for PSR B1957+20. Grey points are taken from \citet{rt91}, and black points from Arecibo observations in 2014 and 2018 (Robert Main priv. comm.). Colours are consistent between panels and represent separate eclipses. \textit{Top}: Full eclipse durations. The fitted power law line has slope $\alpha_{\text{full}}=0.18$ -- fit to all LOFAR points; \textit{Bottom}: Ingress and egress durations. Fitted power law lines have slopes $\alpha_{\text{in}}=0.18$ (solid) -- fit to all LOFAR points; $\alpha_{\text{eg}}=0.21$ (dashed) -- fit to all LOFAR points. Grey and black ingress and egress points have each been offset marginally in frequency for clarity.}
	\label{fig:B1957_pwrlaw}
\end{figure}

In Fig.~\ref{fig:B1957_pwrlaw} the measured eclipse durations for B1957 are shown, along with the power laws fit to our data ($<200$\,MHz). For comparison we also over-plot previously published higher-frequency measurements from \citet{rt91} (grey points), and more recent approximate durations taken from observations with the Arecibo telescope (black points; Robert Main priv. comm.) which were only informally measured, without the use of a model fit. The larger uncertainties on the \citet{rt91} measurements, relative to ours at low-frequency, are primarily a result of temporal variability in the eclipses which has been averaged over, and the plotted errorbars span the regions containing $\sim68$\% of the observed eclipses. This is also the case with the Arecibo data, however the informal `measurement' method led to us assigning conservative uncertainty estimates that reflected our confidence and more than encompassed the full variability between the observed eclipses.\\
\noindent
For the few eclipses at low-frequency that we have analysed, the measured durations are generally consistent with one another, and like PSR J1810+1744, do not display the egress instability that we observe in J1816. Although the higher-frequency points from \citet{rt91} are displaced relative to our LOFAR measurements, we believe that this is likely a consequence of temporal variability or lower signal-to-noise causing a systematic apparent lengthening of the eclipses, and are not representative of the eclipses actually being as long as those at low-frequency. This argument is consistent with the more recent Arecibo data, whereby the eclipse durations $>300$\,MHz are shorter than those $<200$\,MHz, as expected.\\
Taken on their own, ignoring the systematic shift relative to our data, the \textit{egress} measurements $>300$\,MHz suggest a steeper frequency relationship than our measurements $<200$\,MHz. On the other hand, the \textit{ingress} measurements appear more consistent with a single power law component. These are similar trends to those observed for J1816, and Table~\ref{Table: psr_params} shows both that the ingress power laws for both pulsars are consistent within $1\sigma$, and that the egress power laws for the pulsars are significantly closer when fit to the low-frequency data only (Table~\ref{Table: J1816_pwrlaw}). Furthermore, through power law fits to their higher-frequency data, \citet{rt91} measured the full eclipse duration for B1957 to scale with $\alpha \approx 0.4$; entirely consistent with our measured $\alpha_{\text{full}}$ for J1816 when considering the high-frequency data (Table~\ref{Table: psr_params}).

\subsection{PSR J1810+1744}\label{sec: J1810}
\begin{figure}
	\centering
	\includegraphics[width=0.5\columnwidth]{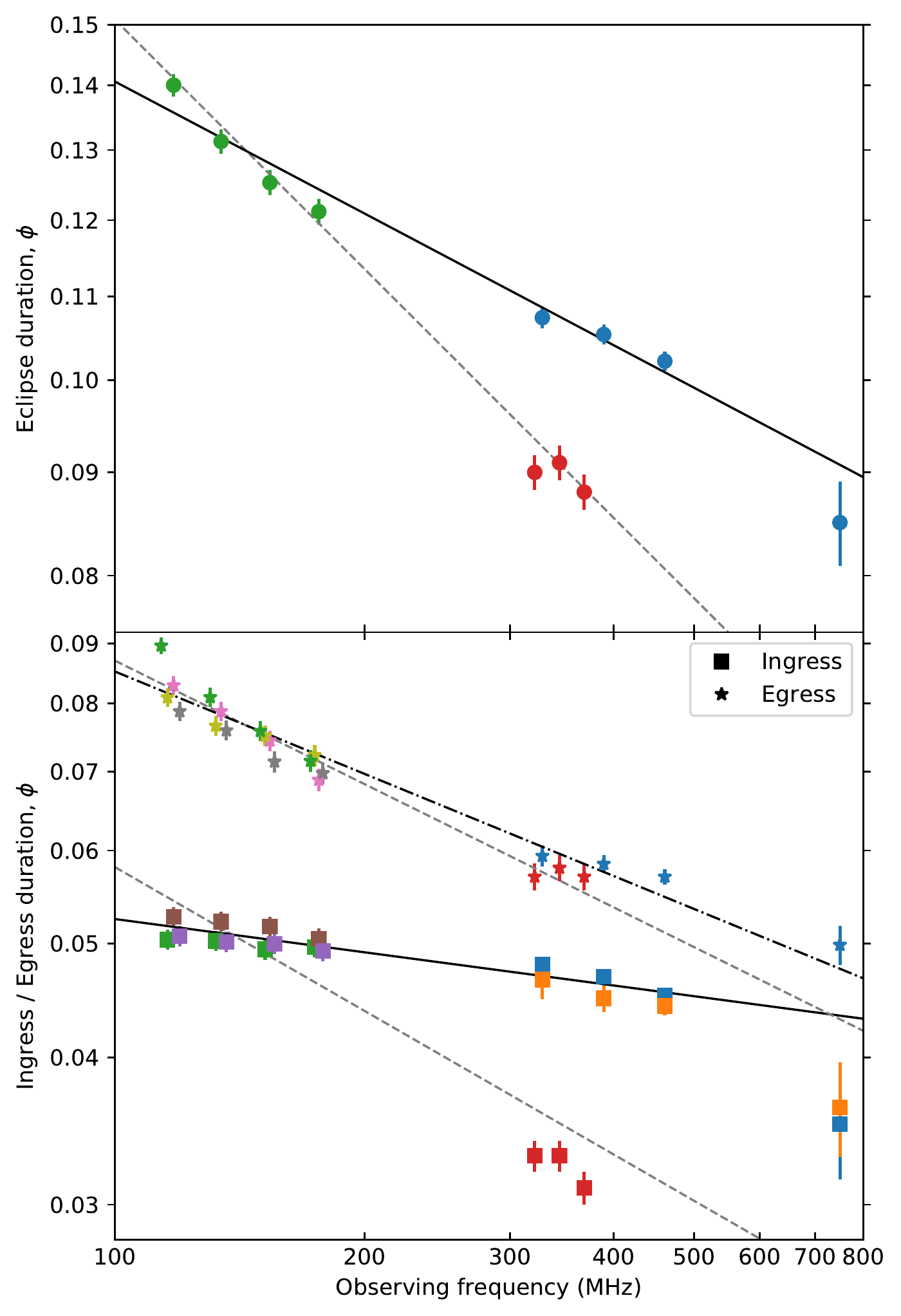}
	\caption[Eclipse durations versus frequency for PSR J1810+1744]{Eclipse durations versus frequency for PSR J1810+1744. Data points $<200$\,MHz, and points in red near 350\,MHz are taken from P18. Colours are consistent between panels and represent separate eclipses. \textit{Top}: Full eclipse durations. Fitted power law lines have slopes $\alpha_{\text{full}}=0.22$ (solid) -- fit to blue GMRT and green LOFAR points; $\alpha_{\text{full}}=0.41$ (grey, dashed, P18) -- fit to red WSRT and green LOFAR points. \textit{Bottom}: Ingress and egress durations. Fitted power law lines have slopes $\alpha_{\text{in}}=0.09$ (solid) -- fit to all LOFAR and blue and orange GMRT points; $\alpha_{\text{in}}=0.41$ (grey, dashed, P18) -- fit to all LOFAR and red WSRT points; $\alpha_{\text{eg}}=0.29$ (black, dash-dot) -- fit to all egress points; $\alpha_{\text{eg}}=0.35$ (grey, dashed, P18) -- fit to all LOFAR and red WSRT points. The poor fit of the 750\,MHz ingress points (blue, orange) suggests a step change in the eclipse medium properties, or a change in eclipse mechanism (see Section \ref{sec: J1810}).}
	\label{fig:J1810_pwrlaw}
\end{figure}

In Fig.~\ref{fig:J1810_pwrlaw} the measured full eclipse, ingress and egress durations for black widow PSR J1810+1744 are plotted, along with the corresponding fitted power laws. Also plotted here are the data from P18, with the fitted power law shown by the grey, dashed line. Large inter-eclipse variability is visible at frequencies near 350\,MHz which appears to be primarily caused by a significant shift in the ingress location, whereas the egress is consistent between the two observed eclipses.\\
\noindent
The WSRT 345\,MHz observation (red points) took place in 2011-Jun, more than a year prior to the earliest 149\,MHz observation (Table~\ref{Table: obs_freq}), and $\sim7$\,yrs prior to the recently observed GMRT eclipse at a similar frequency (blue points), thus the timescale of the ingress shift can only be loosely constrained. We note that there is little variability in the 149\,MHz eclipses over 2012-Dec -- 2015-Feb (although only sparsely sampled). In addition, we obtained 69 single station LOFAR (149\,MHz) observations spread over 2017-Jul -- 2018-Aug. These had too low signal-to-noise to split into sub-bands, but the full bandwidth eclipses were analysed and the vast majority of ingress durations laid in the range $0.052<\Delta\phi_{\text{in}}<0.057$, with extremes at $\Delta\phi_{\text{in}}\approx0.047$ and $\Delta\phi_{\text{in}}\approx0.058$. On the other hand, the majority of egress durations were found to lie within $0.068<\Delta\phi_{\text{eg}}<0.085$, with extremes at $\Delta\phi_{\text{eg}}\approx0.065$ and $\Delta\phi_{\text{eg}}\approx0.095$, demonstrating much larger variability in the eclipse egress relative to ingress at these low-frequencies. The high signal-to-noise LOFAR Core observations plotted in Fig.~\ref{fig:J1810_pwrlaw} are consistent with the most common range of the single station observations for both ingress and egress.\\
Taking into account the implied higher variability in the eclipse egress from the LOFAR single station observations, and also noting the consistency of the plotted 149\,MHz and GMRT 400\,MHz ingress durations with a single power law -- reduced $\chi^2 = 1.6$ as opposed to a reduced $\chi^2 = 8.6$ for the LOFAR and WSRT fit from P18 -- suggests that WSRT observation may have caught a relatively rare event, or that the higher density inner material is more variable than the tenuous outer material responsible for 149\,MHz eclipse boundaries. In either case, we believe that the new fitted power laws presented here give a more reliable representation of the system over this range of frequencies.\\
Moving to higher frequencies, the eclipse simultaneously observed at 400\,MHz and 750\,MHz interestingly does not appear to follow a simple power law -- although the uncertainty is relatively large on the 750\,MHz observations -- which could be a result of the eclipse medium not being smoothly varying over this range, or possibly different eclipse mechanisms. Although we observed two consecutive eclipses at these frequencies, only one eclipse egress is plotted due to unusual behaviour in the second egress, whereby the flux density at both 400\,MHz and 750\,MHz initially began to re-emerge at the same phase as the previous eclipse, then sharply disappeared for the remainder of the observation -- only a few minutes. Such `mini-eclipses' with durations on the order of a few minutes, close to eclipse edges, were also observed in the BW pulsar PSR J1544+4937, and were attributed to clumps of plasma surrounding the main eclipse material \citep{brr+13}. In cases where the flux is not detected to briefly re-emerge, these clumps could be responsible for the variable eclipse durations that we observe.

\subsection{PSR J2051$-$0827}\label{sec: J2051}
\begin{figure}
	\centering
	\includegraphics[width=0.5\columnwidth]{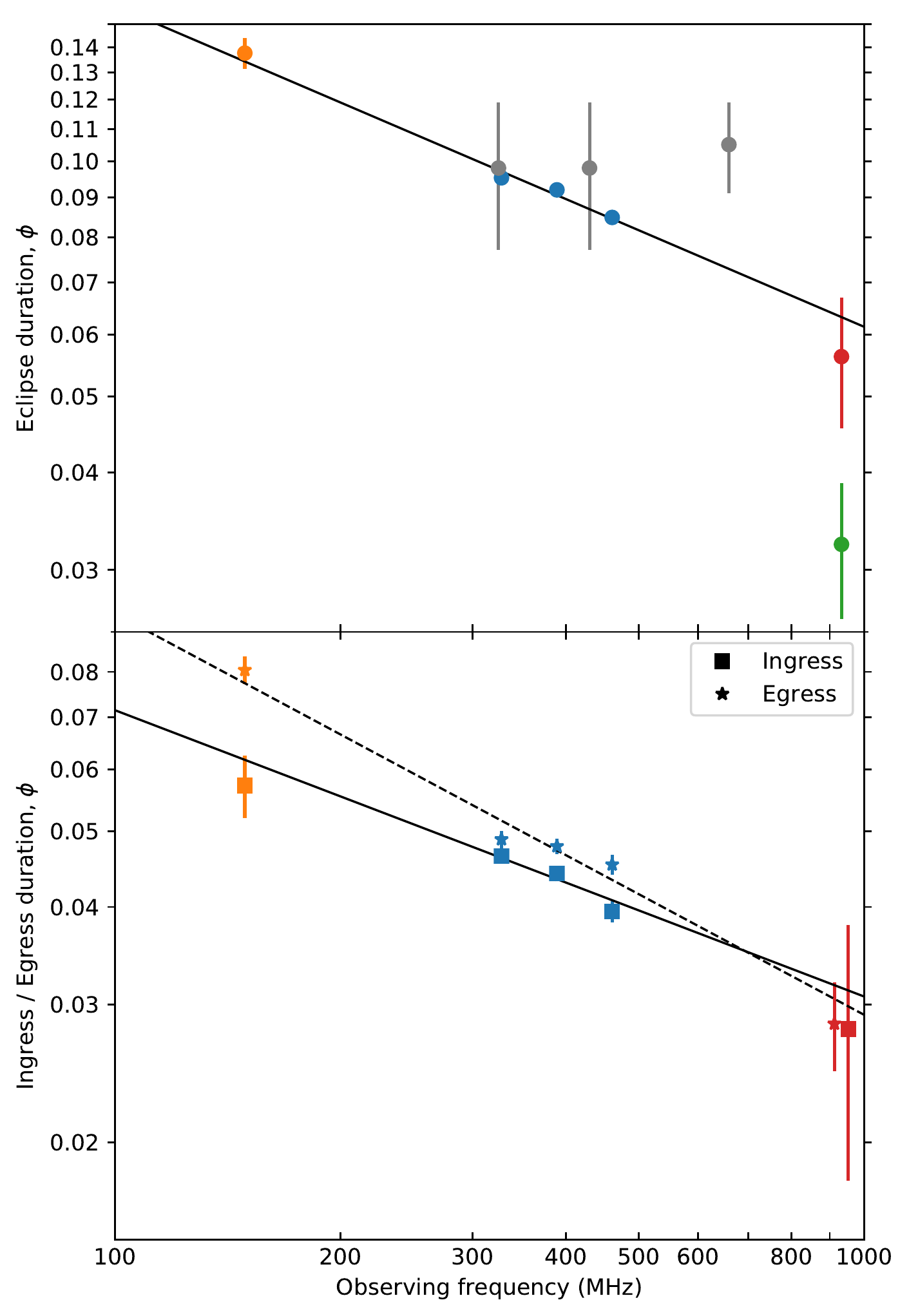}
	\caption[Eclipse durations versus frequency for PSR J2051$-$0827]{Eclipse durations versus frequency for PSR J2051$-$0827. Grey points are taken from \citet{sbl+01}. Colours are consistent between panels and represent separate eclipses. \textit{Top}: Full eclipse durations. The fitted power law line has slope $\alpha_{\text{full}}=0.41$ -- fit to all LOFAR and GMRT points and red Parkes point; \textit{Bottom}: Ingress and egress durations. Fitted power law lines have slopes $\alpha_{\text{in}}=0.37$ (solid) -- fit to all ingress points; $\alpha_{\text{eg}}=0.51$ (dashed) -- fit to all egress points. Red Parkes ingress and egress points have each been offset marginally in frequency for clarity.}
	\label{fig:J2051_pwrlaw}
\end{figure}

\begin{figure*}
	\includegraphics[width=.9\textwidth]{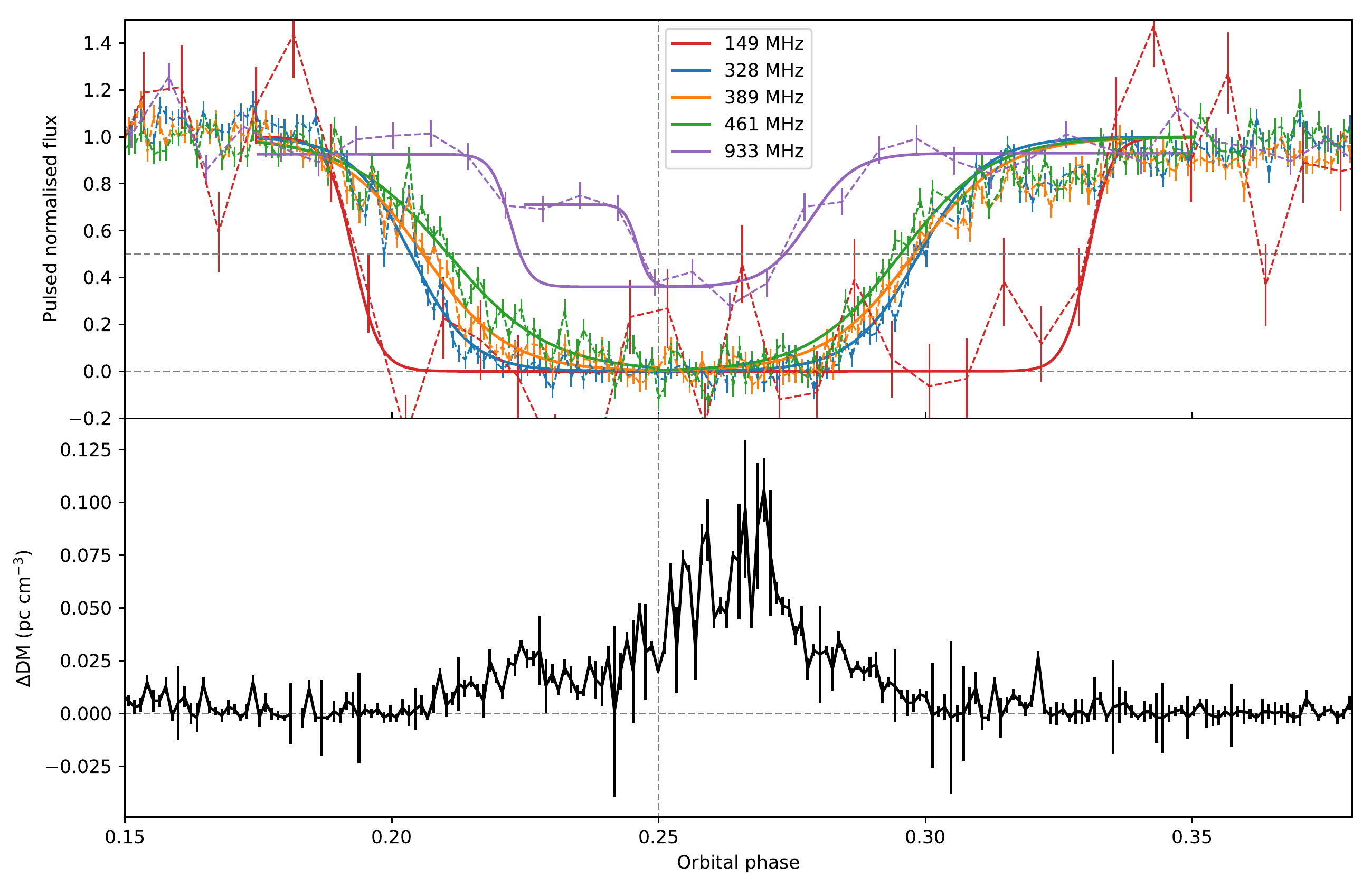}
	\caption[Eclipse model fitting for PSR J2051$-$0827]{\textit{Top}: Measured pulsed flux densities (dashed) for PSR J2051$-$0827, normalised so that the mean out-of-eclipse levels are unity, taken from P19. Solid lines show the best-fit Fermi-Dirac functions for each ingress and egress (see Section \ref{sec: analysis_ecl_dur}). The half-flux density level is represented by the thin horizontal dashed line. \textit{Bottom}: DM relative to mean out-of-eclipse value measured from a single eclipse observation spanning 705--4032\,MHz (P19).}
	\label{fig:J2051_fits}
\end{figure*}

PSR J2051$-$0827 is a well studied pulsar and has been observed over a wide range of frequencies for multiple decades (e.g. P19). However, its relatively short duration eclipses mean that even small amounts of temporal variability can completely mask any genuine dependence of the eclipse duration on frequency, as was found to be the case in the only previous attempt to measure the frequency dependence \citep{sbl+01}. To try and overcome this problem, we undertook a campaign to perform coordinated multi-frequency observations of the eclipses over just a two week period. The results of these observations are shown in Fig.~\ref{fig:J2051_pwrlaw}, along with the power law curves fit to our data. Over-plotted in grey are the measured durations at 325\,MHz, 430\,MHz and 660\,MHz taken from \citet{sbl+01}, with the uncertainties representing the variability observed in 7, 22 and 10 eclipses, respectively, over the dates 1994-May -- 1997-Sep.\\
\noindent
We note that the `eclipse' at 933\,MHz did not fully remove the pulsed flux density, and instead only reduced it to $\sim40$\% of the out-of-eclipse level. The eclipse was also somewhat irregular, appearing to occur in two stages, with an initial drop down to $\sim70\%$ of the out-of-eclipse level where it remained in an intermediate plateau, before dropping a second time; conversely, the egress appeared to be more regular. In an attempt to model this eclipse we fitted two separate ingress functions, with the `outer-eclipse' fit to the initial flux decrease, while masking the plateau data points, and the 'inner-eclipse' fit to the second flux decrease, taking the plateau as the effective out-of-eclipse level. The observed eclipses are shown in the top panel of Fig.~\ref{fig:J2051_fits}, over-plotted with the fitted model functions as described in Section~\ref{sec: analysis_ecl_dur}.\\
\noindent
In the top panel of Fig.~\ref{fig:J2051_pwrlaw} we plot the effective eclipse durations for both the inner- (green) and outer-eclipses (red) at 933\,MHz, showing only the outer-eclipse to be consistent with a single power law across all the measured frequencies. Irrespective of this observation, the lower-frequency data show a clear frequency dependence of the eclipse duration, allowing us to successfully measure a power law relationship for this pulsar for the first time.\\
Comparing the eclipse durations from \citet{sbl+01} with ours, it can be seen that at 325\,MHz and 430\,MHz both sets are consistent, however their duration at 660\,MHz is far longer than predicted by our power law fit. To investigate this one can refer to the DM measurements across the eclipse region, which are representative of the material density distribution. In the bottom panel of Fig.~\ref{fig:J2051_fits} the measured $\Delta$DM profile corresponding to the time of the observing campaign is shown. The profile is asymmetric about companion inferior conjunction, with a brief peak in density ($\Delta\text{DM} > 0.05$\,pc\,cm$^{-3}$) occurring only over the phase range $\approx0.25$--0.275, coincident with the 933\,MHz inner-eclipse. Conversely, in Fig.~6 of \citet{sbl+01} their measured DM profile has a peak that, although of a similar order of magnitude, persists for much longer, covering $\phi \approx 0.22$--0.29. If we assume this to be representative of the DM profile for most of their observed 660\,MHz eclipses -- in P19 similar DM structures are shown to persist for months-to-years -- then this suggests that the higher-frequency eclipses are sensitive only to the most dense regions of the eclipse material, which is shown in P19 to be highly variable over multiple years. On the other hand, the low-frequency eclipses appear to be more stable (P19) and less sensitive to the peaks of the density profile close to the companion, which they do not directly probe. This highlights the susceptibility of eclipse duration measurements to become erratic in the presence of irregular material density distributions. With the low inclination angle of the orbit of PSR J2051$-$0827 implying that our line of sight cuts across the outer edge of the ablated material, this pulsar in particular may be at risk of asymmetric and irregular density distributions, which is also implied by the work in P19.

\subsection{PSR J2215+5135}\label{sec: J2215}
\begin{figure}
	\centering
	\includegraphics[width=0.5\columnwidth]{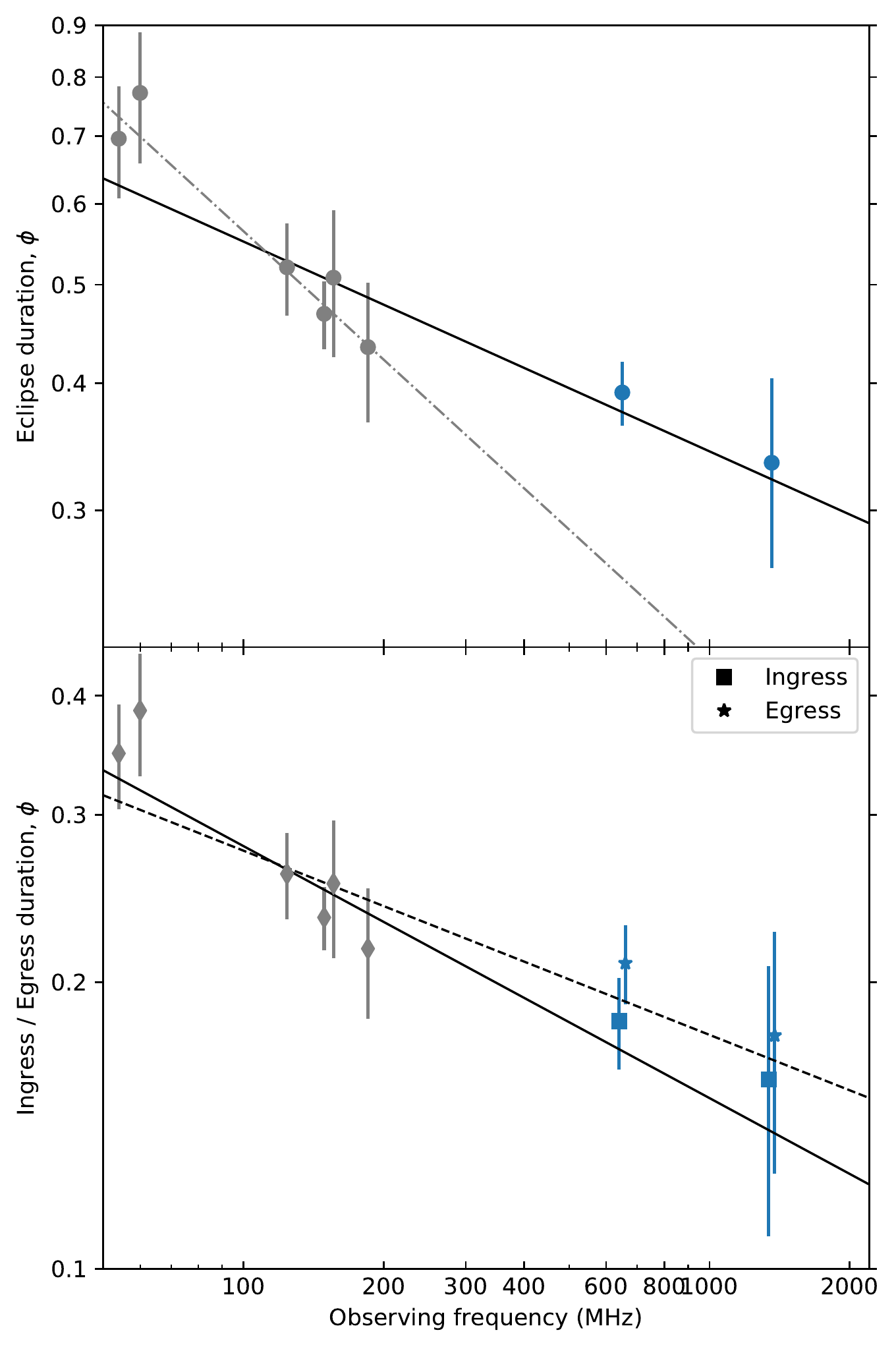}
	\caption[Eclipse durations versus frequency for PSR J1810+1744]{Eclipse durations versus frequency for PSR J2215+5135. Grey points $<200$\,MHz are taken from \citet{bfb+16}. Colours are consistent between panels and represent separate eclipses. \textit{Top}: Full eclipse durations. Fitted power law lines have slopes $\alpha_{\text{full}}=0.21$ -- fit to all points; $\alpha_{\text{full}}=0.42$ \citep[grey, dash-dot,][]{bfb+16} -- fit to LOFAR points only; \textit{Bottom}: Ingress and egress durations, grey points $<200$\,MHz represent equal ingress and egress durations as \citet{bfb+16} assumed symmetric eclipses. Fitted power law lines have slopes $\alpha_{\text{in}}=0.26$ (solid) -- fit to all ingress points; $\alpha_{\text{eg}}=0.19$ (dashed) -- fit to all egress points. Blue GMRT ingress and egress points have each been offset marginally in frequency for clarity.}
	\label{fig:J2215_pwrlaw}
\end{figure}

For the final pulsar in the study, we have observed a single eclipse simultaneously in two relatively high frequency bands. The measured eclipse durations are shown in Fig.~\ref{fig:J2215_pwrlaw}, along with low-frequency data from LOFAR imaging observations published in \citet{bfb+16}. The black power law curves resulted from fits to all of the available data, \textit{including} that of \citet{bfb+16}, while the grey, dash-dot line shows the published power law fit to the LOFAR data only. The errorbars assigned to our measurements have been increased from the formal $1\sigma$ fits in order to better reflect our confidence in the model fits due to the presence of large gaps ($\Delta\phi \sim 0.04$) in the data masking the beginning of both the ingress and egress, at times when phase calibration of GMRT antennas took place.\\
\noindent
It can be seen that the eclipses in PSR J2215+5135 cover huge fractions of the orbit relative to the previous pulsars in this study. This is typical of RB pulsars, and especially those that are known to transition between RB states and accretion states \citep[e.g.][]{asr+09}. These long eclipse durations mean that their frequency dependence is generally easier to detect, even though the power law fit to our data is relatively shallow (see Table~\ref{Table: psr_params}). The implied frequency dependence, with $\alpha_{\text{full}}=0.21\pm0.04$, is at odds with the significantly steeper one measured across frequencies $<200$\,MHz, with $\alpha=0.42\pm0.11$ \citep{bfb+16}, however is still consistent with all of the data (reduced $\chi^2 = 0.8$). It is worth noting that as we have only observed a single eclipse at higher frequencies, there is no information on the magnitude of temporal variability in the eclipses, and this may play a role in the discrepancy between the two measured power law exponents.

\section{Discussion}\label{sec: discuss}
As an indication of the volume of excess material in and around the orbits of J1816 and B1957 we can draw comparisons between the $\Delta$DM measurements. Beginning with the trailing tail of material post-eclipse, our LOFAR observations show $\Delta\text{DM} \sim 0.01$\,pc\,cm$^{-3}$ for both near orbital phase $\phi = 0.32$. Probing closer to the companion, the 650\,MHz observation of J1816 shows $\Delta$DM is around a factor of 10 larger than the above LOFAR measurements by $\phi=0.275$. Similarly, the observations of B1957 presented by \citet{rt91} show TOA delays of $\sim200$\,$\mu$s at 1400\,MHz at this orbital phase, also corresponding to $\Delta\text{DM} \sim 0.1$\,pc\,cm$^{-3}$. Prior to the eclipse, in J1816 the ingress $\Delta$DM appears to be negligible until slightly closer to companion inferior conjunction relative to B1957, although the TOA measurements in \citet{slr+14} suggest a sharp increase in DM at ingress, just beyond the beginning of the eclipses at 650\,MHz observed here, again comparable with that in B1957 \citep{rt91}.\\
Considering that the orbital separations for both systems are much the same (see Table~\ref{Table: psr_params}), and assuming that the material resides in close proximity to the orbits, then the implied physical widths spanned by the material are similar. Taking the common approximation that the material depth is equivalent to the eclipse width, then the implied density profiles for the two systems are also similar.\\
The comparable eclipse material densities, volumes and therefore masses that are inferred from the $\Delta$DM measurements alone has interesting connotations when the vast differences between the nature of the companion stars are considered. On the one hand, these observations may arise out of a coincidence in the balance of competing effects caused by differences in the companion stars other than just their masses, e.g. density of the star and its size relative to the Roche lobe, strength of a stellar magnetic field etc., which may result in similar amounts of ablated material. With this in mind, we note that if both companions fill their respective Roche lobes, then the implied total pulsar wind energies incident on the stars are approximately equal, i.e. $U_{E,\text{B}1957} R_{L,\text{B}1957}^2 \approx U_{E,\text{J}1816} R_{L,\text{J}1816}^2$ (Table~\ref{Table: psr_params}), in the case of isotropic winds.\\
Alternatively, if the eclipse material instead originates from electron-positron pairs in the pulsar wind becoming trapped in a magnetosphere \citep[e.g.][]{kmg00,llm+19}, then the magnetic fields of the companions would dictate the mass and volume of eclipse material rather than the companion masses. However, a study into the radio emission from B1957 presented in \citet{llm+19} appears to cast doubt on the latter scenario, as the observed polarisation properties did not agree with those theoretically predicted for propagation of the radio emission through a magnetised pair plasma. The authors do directly state that this could still be a promising scenario for J1816, for which ablation of material from the companion would be difficult to explain given its expected degenerate nature \citep{kbv+13}. Irrespective of its origin, the suggested similarities between the material around the companion stars in these systems means that J1816 would make an interesting candidate for lensing of pulsar radio emission pre- and post-eclipse by the clumpy material, as recently observed in B1957 \citep{myc+18}.\\
Regarding the timing irregularities in B1957 presented in Section~\ref{sec: B1957_toas}, neither the source nor the mechanism are clear, though broadening of the tail of pulses is indicative of scattering and/or dispersion. The precision with which we can measure TOAs in J1816 is not as high -- due to the lower flux density of the pulsar -- however we see no evidence for structure of a similar order of magnitude in the TOA residuals. In addition, there are no clear indications of such variations in our data for PSRs J1810+1744, J2051$-$0827 or J2215+5135. Further long duration observations of B1957, covering all orbital phases, would provide the opportunity to test for correlation between the TOA variations and orbital phase, which, if found, would strongly suggest that the origin lies within the binary system, and likely a result of propagation effects in excess material.\\
The globular cluster eclipsing pulsar, PSR B1744$-$24A (Ter5A), is a well studied example of a binary system which demonstrates propagation delays that are caused by the ablated material from the companion unusually overcoming the pulsar wind, and flowing all around the orbit \citep{lmd+90}. The propagation effects are much more significant in Ter5A, and are often accompanied by eclipses of variable durations occurring at all orbital phases, however it may be the case that the outflow geometry in B1957 bears some similarities to Ter5A, albeit with less material flowing around the orbit. Indeed, \citet{tb91,tb93} model the outflows in both B1957 and Ter5A using two-dimensional hydrodynamical simulations, and by balancing their model parameters -- mass loss rate, gas injection temperature, Mach number, effective cross-section and pulsar luminosity -- can qualitatively reproduce the basic observable properties of the two. There is significant flexibility in these model parameters as they are largely unconstrained by independent observations, thus it would be interesting to investigate if realistic tweaking of the parameters could allow some of the ablated material in B1957 to also flow around the orbit.

\subsection{Mass loss}\label{sec:mass_loss}
The studies in P18 and P19 made use of the method presented in \citet{t+94} in order to estimate mass loss rates from spider systems. This assumes that the material leaves the system through a projected circle of diameter equal to the eclipse width, perpendicular to the orbital plane. The velocity of the material is determined by equating the energy density of the pulsar wind at the companion distance to the momentum flux of the eclipse material, i.e. $V_{\text{W}} = (U_{\text{E}}/n_{\text{e}}m_{\text{p}})^{1/2}$. Then the mass loss rate is given by $\dot{M}_{\text{C}} \sim \pi R_{\text{E}}^2m_{\text{p}}n_{\text{e}}V_{\text{W}}$, where $R_{\text{E}}$ is the half-width of the eclipse.\\
In \citet{t+94} the authors specifically focussed on the case of B1957, estimating $\dot{M}_{\text{C}} \sim 3\times10^{-13}~M_\odot$~yr$^{-1}$. For consistency we can make a new approximation for this based on the 149\,MHz observations, and taking into account the expected orbital inclination. Using the values in Table~\ref{Table: psr_params} the equivalent eclipse width at the companion's orbit is $2 \pi a_C \phi_{\text{ecl}} \approx 1.64 R_{\odot}$, for $a_C = a_p M_{\text{PSR}} / M_C$, $a_p \sin i = 0.089$\,lt-s \citep{aft94}, $M_{\text{PSR}} = 2.4 M_{\odot}$ and $M_C = 0.035 M_{\odot}$ \citep{vbk11}. When viewed at an inclination of $65^{\circ}$ this corresponds to a chord across the assumed circle with diameter $2.98 R_{\odot}$. Taking the 149\,MHz $\Delta\text{DM} = 0.01$\,pc\,cm$^{-3}$ to be distributed over a depth similar to the eclipse radius gives an implied density, $n_{\text{e}} = 1.5 \times 10^5$\,cm$^{-3}$. This gives an expected mass loss rate, $\dot{M}_{\text{C}} \sim 10^{-12}~M_\odot$~yr$^{-1}$.\\
For J1816 the inclination of the orbit is not yet constrained \citep{kbv+13}, and as such we assume here that $i = 90^{\circ}$. Again using values from Table~\ref{Table: psr_params}, and an eclipse duration of 11\% of the orbit, the estimated eclipse width is $2 \pi a_C \phi_{\text{ecl}} \approx 1.70 R_{\odot}$, for $a_C = a_p M_{\text{PSR}} / M_C$, $a_p \sin i = 0.60$\,lt-s \citep{slr+14}, $M_{\text{PSR}} = 1.84 M_{\odot}$ and $M_C = 0.193 M_{\odot}$ \citep{kbv+13}. For $\Delta\text{DM} = 0.01$\,pc\,cm$^{-3}$, the implied density of eclipse material is $n_{\text{e}} = 2.6 \times 10^5$\,cm$^{-3}$ if the depth is taken to be comparable to the eclipse width. This gives an estimated mass loss rate for J1816 of $\dot{M}_{\text{C}} \sim 2 \times 10^{-13}~M_\odot$~yr$^{-1}$.\\
These estimated values are of a similar order of magnitude to those found for black widow pulsars PSR J1810+1744 and PSR J2051$-$0827 in P18 and P19, respectively, and the same caveats apply, in that the initial assumptions of the mass loss process heavily dominate the estimated values. Considerations such as a companion magnetosphere trapping the material, or widening of the orbit as mass is lost could lead to much lower, or continually decreasing, mass loss rates. On the other hand, if the material has a large neutral component then the mass loss could be much larger, as the DM measurements are only sensitive to ionised material. Assuming that the latter case is unlikely, bearing in mind the expected high energies incident upon the material \citep{t+94}, it would appear that these estimates are more appropriately considered as upper limits, suggesting that complete evaporation of the companion stars does not appear realistic within typical timescales of evolution, especially for the more massive companion of J1816.\\
The surmounting evidence against observed spider pulsars having high enough mass loss rates to completely evaporate their companion stars leads to interesting questions regarding their evolution. An increasing discovery rate of eclipsing spiders, with similar companion star masses, naively suggests that these systems can persist in a relatively stable state for long time scales, with only negligible mass loss from the companion. What happens to these systems in the long-run remains unanswered. We note recent work by \citet{gq20} in which they model outflows from black widow companions considering thermal and dynamical processes in the wind. They estimate the companion of B1957 to be evaporated in 3.4\,Gyr, around a factor of 10 faster than our naive estimates above, however they reach the same conclusion in that evaporation alone appears unlikely to be able to explain a direct link between isolated MSPs and black widows. Alternatively, the authors suggest that a magnetic braking mechanism, whereby the pulsar wind couples to the companion magnetic field, can produce estimated black widow companion lifetimes of a few -- tens of Gyr and argue that this is more consistent with the observed populations of black widows and isolated MSPs.

\subsection{Mechanisms}\label{sec:mechanisms}
As the `original' black widow, B1957 has been subject to many in-depth scrutinies, and the mechanisms behind the observed eclipses have been of particular interest. \citet{t+94} reviewed many of the previously proposed mechanisms, and also examined some seemingly promising alternatives. The authors found smearing out of the pulsations to be most consistent with the observational evidence for eclipses at 1400\,MHz, and suggested that cyclotron-synchrotron absorption in either a companion star magnetosphere, or material entrained in the pulsar wind, was most likely at lower-frequencies. However, nonlinear Raman scattering also provided a feasible alternative at low-frequencies, should the plasma conditions allow growth of induced turbulence to a magnitude large enough to remove pulsar flux from the line of sight. Later, \citet{kmg00} presented a different model of cyclotron damping of the radiation, considering a companion magnetosphere filled with electron-positron pairs from the pulsar wind, which was again consistent with the available observations.\\
Recently some difficulties have arisen for both of the proposed cyclotron models, as new observations of lensing, and polarisation, in B1957 have led to tight upper limits on the strength of magnetic fields near the eclipse edges \citep[$B_{||} = 0.02 \pm 0.09$\,G;][]{llm+19}, which are seemingly lower than those required for either model. Although this constraint is many magnitudes more precise than the limit found for PSR J2051$-$0827 in P19 ($B_{||} = 20 \pm 120$\,G), their observations do not strongly constrain the field strength actually \textit{within} the eclipses where the absorption occurs. No such constraints are available for J1816 due to its low fractional polarisation in the pulsed radio emission.\\
Our observations conclusively show that flux is removed from the line of sight during the low-frequency eclipses of both pulsars, consistent with absorption or nonlinear scattering eclipse mechanisms suggested above. In addition, in Section~\ref{sec: J1816} we show that the frequency dependence of the duration of eclipses in J1816 is consistent with the $\nu^{-0.4}$ relation found for B1957 over a wide range of frequencies \citep{fbb+90,rt91}, suggesting that a similar mechanism could be responsible for both.\\
For J1816, and to a lesser extent B1957, our observations show that the pulsed emission can sometimes remain undetected in the variable tail of material, while the continuum flux density recovers to the out-of-eclipse level, strongly suggesting that the propagation through the material causes the pulsations to become smeared in time. As considered in both P18 and P19, rapid fluctuations in DM can smear out pulsations if the change in induced delays within a single time-integration significantly exceeds the pulse width. For J1816, with a pulse width approximately half of the 3.19\,ms pulse period, the necessary change in DM within an integration would need to exceed $\sim0.01$\,pc\,cm$^{-3}$ at 149\,MHz. The 650\,MHz observation offers a probe into the smearing region, and shows the DM variation to in fact be relatively slow, with the gradient (resolved over 10\,s integrations at 650\,MHz) to be often significantly below that required to smear out the 149\,MHz pulsations within a 30\,s integration. The required change in DM within a sub-integration is shown in Fig.~\ref{fig:J1816_flux_dm} with a black line for comparison, and the equivalent for B1957 in Fig.~\ref{fig:B1957_flux_dm}.\\
An alternative to DM smearing regards an increase in scattering causing the pulsations to become broadened in time. As the 149\,MHz pulsed flux density recovers immediately after the smearing region, our fits of $\Delta$DM and $\Delta\tau$ are consistent with a rise in scattering timescale relative to the out-of-eclipse level, but this is loosely constrained to be $\lesssim 0.2P$, where $P$ is the pulse period. As such, for scattering to cause the loss of pulsations, a relatively sharp boundary must be present where $\Delta\tau$ increases to $>0.5P$ within a 30\,s integration. Again utilising the 650\,MHz observation to directly probe the smearing region, our model fits constrain $\Delta\tau \lesssim 0.03P$ at 650\,MHz. Taking the scattering timescale to scale with frequency as $\tau \propto \nu^{-4}$ \citep{l71,lj76}, this corresponds to roughly 10 times the pulse period at 149\,MHz, and so is entirely consistent with scatter-broadening causing the loss of pulsations during egress.\\
Should scattering be the root cause of the variable J1816 egresses, this implies differences in the tail material of the two pulsar systems that is not noticeable from the DM variations alone. As B1957 has a similar pulse profile to J1816, but with a pulse period approximately half as long, the threshold scattering timescale to smear out the pulsations would also halve. As a consequence, if the properties of the tail material were to be similar in both systems, the pulsations in B1957 should also undergo smearing beyond the pulse period. Although there is some evidence of pulse smearing in the B1957 egresses, this is to a much lesser extent than that seen for J1816. Such differences may be attributable to the inclination with which we view the systems, whereby the general material properties (turbulence, clumping etc.) may change with radial distance from the star. Without further knowledge of the inclination of the J1816 orbit we pose this as open question.

\subsection{Frequency dependence}\label{sec: disc_freq}
In Table~\ref{Table: psr_params} we show the measured frequency power law exponents along with the properties of the observed binary systems. Also included are previously published measurements for Ter5A, whereby \citet{ntt+90} found an apparently steep eclipse duration dependence on frequency of $\Delta\phi \propto \nu^{-0.63\pm0.18}$.\\
\begin{table*}
	\centering
	\caption[Orbital parameters and frequency dependence of eclipses]{Parameters of spider systems along with best-fit power law exponents for eclipse full-width half-maxima, $\alpha_{\text{full}}$, ingress duration, $\alpha_{\text{in}}$, and egress duration, $\alpha_{\text{eg}}$, against observation frequency, i.e. $\Delta\phi_{\text{eclipse}} \propto \nu^{-\alpha}$. Three $\alpha_{\text{full}}$ and $\alpha_{\text{eg}}$ are listed for PSR J1816+4510 due to the observed high eclipse-to-eclipse variability in the egress phase. The pulsar binary parameters are: companion mass, $M_C$, and radius, $R_C$, approximate Roche lobe radius, $R_L$, orbital period, $P_b$, orbital separation, $a$ -- calculated from published projected semi-major axis, $a_p\text{sin}i$, orbit inclination, $i$, and mass ratio as $a = a_p + a_p M_{\text{PSR}} / M_C$ -- and energy density of pulsar wind at the companion distance, $U_E = \dot{E} / 4\pi ca^2$. Also included for comparison are the parameters for globular cluster PSR B1744$-$24A (Ter5A), not measured in this work. $^{\text{a}}$~\citet{bvr+13}; $^{\text{b}}$~Minimum mass assuming $M_{\text{PSR}} = 1.4M_{\odot}$ and $i=90^{\circ}$; $^{\text{c}}$~Assuming 80\% Roche lobe filling factor \citep{bvr+13}; $^{\text{d}}$~\citet{hrm+11}; $^{\text{e}}$~\citet{kbv+13}; $^{\text{f}}$~\citet{slr+14}; $^{\text{g}}$~\citet{vbk11}; $^{\text{h}}$~\citet{aft94}; $^{\text{i}}$~\citet{t+94}; $^{\text{j}}$~\citet{lvt+11}; $^{\text{k}}$~\citet{svb+01}; $^{\text{l}}$~\citet{svf+16}; $^{\text{m}}$~\citet{lsc18}, assuming 95\% Roche lobe filling factor; $^{\text{n}}$~\citet{aaa+13}; $^{\text{p}}$~Optical observations by \citet{lsc18} suggest that this could be a factor of 3 larger; $^{\text{q}}$~Fit included LOFAR data from \citet{bfb+16}; $^{\text{r}}$~\citet{nt92}; $^{\text{s}}$~Approximate Roche lobe radius assuming $i=90^{\circ}$ \citep{nt92}; $^{\text{t}}$~\citet{ntt+90}.}
	\label{Table: psr_params}
	\begin{tabular}{lcccccccccc}
		\hline	
		PSR & $M_C$ & $R_C$ & $R_L$ & $P_b$ & $a$ & $i$ & $U_E$ & $\alpha_{\text{full}}$ & $\alpha_{\text{in}}$ & $\alpha_{\text{eg}}$ \\
		& ($M_{\odot}$) & ($R_{\odot}$) & ($R_{\odot}$) & (h) & ($R_{\odot}$) &  & (ergs\,cm$^{-3}$) &  &  &  \\
		\hline
		J1810+1744 & 0.045$^{\text{a,b}}$ & 0.15$^{\text{a,c}}$ & 0.15$^{\text{a}}$ & 3.6$^{\text{d}}$ & 1.33$^{\text{a}}$ & $50^{\circ}$$^{\text{a}}$ & 12.3$^{\text{a}}$ & $0.22\pm0.02$ & $0.09\pm0.01$ & $0.29\pm0.02$ \\
		J1816+4510 & $\left(\frac{0.193}{\sin^3 i}\right)$ $^{\text{e}}$ & $\geq 0.26$$^{\text{e}}$ & $0.56$$^{\text{e}}$ & 8.7 & 2.7$^{\text{e,f}}$ & $\leq 90^{\circ}$$^{\text{e}}$ & 3.8$^{\text{f}}$ & $0.49\pm0.06$ & $0.16\pm0.01$ & $0.68\pm0.11$ \\
		& & & & & & & & $0.42\pm0.04$ &  & $0.58\pm0.07$ \\
		& & & & & & & & $0.34\pm0.02$ &  & $0.46\pm0.02$ \\
		B1957+20 & 0.035$^{\text{g}}$ & 0.25$^{\text{g}}$ & 0.29$^{\text{g}}$ & 9.2$^{\text{h}}$ & 2.9$^{\text{g,h}}$ & $65^{\circ}$$^{\text{g}}$ & 14$^{\text{i}}$ & $0.18\pm0.04$ & $0.18\pm0.04$ & $0.21\pm0.09$ \\
		J2051$-$0827 & 0.05$^{\text{j}}$ & 0.064, & 0.15$^{\text{j}}$ & 2.4$^{\text{j}}$ & 1.12$^{\text{l}}$ & $40^{\circ}$$^{\text{k}}$ & 2.5$^{\text{l}}$ & $0.41\pm0.04$ & $0.37\pm0.06$ & $0.51\pm0.07$ \\
		& & 0.14$^{\text{j,k}}$ & & & & & &  &  &  \\
		J2215+5135 & 0.33$^{\text{m}}$ & 0.39$^{\text{m}}$ & 0.41$^{\text{m}}$ & 4.2$^{\text{d}}$ & 1.77$^{\text{m,n}}$ & $63.9^{\circ}$$^{\text{m}}$ & 9.3$^{\text{a,p}}$ & $0.21\pm0.04$$^{\text{q}}$ & $0.26\pm0.04$$^{\text{q}}$ & $0.19\pm0.05$$^{\text{q}}$ \\
		Ter5A & $\geq 0.089$$^{\text{r}}$ & 0.15$^{\text{s}}$ & 0.15$^{\text{s}}$ & 1.8$^{\text{r}}$ & 0.85$^{\text{r}}$ & $\leq 90^{\circ}$ & $2.0\left(\frac{\dot{P}}{1\times10^{-19}}\right)I_{\text{PSR},45}$$^{\text{i}}$ & $0.63\pm0.18$$^{\text{t}}$ &  & \\
		\hline
	\end{tabular}
\end{table*}\noindent
All of the pulsars studied here exhibit a clear frequency dependence of the eclipse duration, with the measured full eclipse duration exponents spread over $0.18 \lesssim \alpha_{\text{full}} \lesssim 0.63$, ingress durations between $0.09 \lesssim \alpha_{\text{in}} \lesssim 0.37$ and egress durations between $0.19 \lesssim \alpha_{\text{eg}} \lesssim 0.68$. For two of the pulsars -- J1816 and B1957 -- a single power law does not appear sufficient to model the eclipse egress durations across both frequencies $<200$\,MHz and $>200$\,MHz, which we interpret as a consequence of a tenuous swept-back tail of material \citep{fbb+90} that has less influence on the propagation of higher-frequency emission due to its likely low-density. This could cause a break in the power law if the eclipse mechanism in the tail material differed to that caused by the main bulk of the medium, or if the properties of the tail material vary with orbital phase at a different rate relative to the main bulk of the medium. For those pulsars where we have wide frequency coverage, namely, PSR J1810+1744, J1816 and PSR J2051$-$0827, the eclipses become continuously more symmetric about companion inferior conjunction towards higher observing frequencies. This is also seen in B1957 if we consider the higher-frequency measurements from \citet{rt91} along with our own. This is further consistent with the idea of a tenuous swept-back tail of material, having increasing influence on the wave propagation towards lower-frequencies.\\
Comparing the properties of the pulsars' orbits with their respective $\alpha$ values, we see no clear correlation with any individual parameter, although there is a hint of a trend of \textit{decreasing} steepness of the power law for \textit{increasing} pulsar wind energy density at the companion star, i.e. $\alpha(U_E)$. However with only 6 samples, and possibly inaccurate or time-dependent estimates of $U_E$ and $\alpha$, this is currently inconclusive. Future optical observations can better constrain the pulsar wind flux incident on the companion stars \citep[e.g.][]{lsc18} and further measurements of the eclipse durations can provide information on the magnitude of temporal variability, thus paving the way to being able to robustly determine the presence of correlations.\\
Some of the differences between the measured frequency dependencies may be due to different eclipse mechanisms, for example \citet{t+94} find the predictions of stimulated Raman scattering to best match the observed eclipse properties in Ter5A, whereas absorption is favoured in B1957. However, equally important are the different electron density distributions in the eclipse media between systems. The influence of different electron distributions will depend on both the properties of the material as a whole and the particular lines of sight that we sample, given the different orbit inclinations. Notably, in PSR J2051$-$0827 we appear to be sampling the `upper-edge' of the material \citep[P19;][]{svb+01} with its relatively low inclination angle, thus the density distribution may be significantly different here to that crossed by a line of sight passing closer to the companion. Indeed, the DM measurements throughout the eclipse region in PSR J2051$-$0827, presented in P19, often show that stochastic clumpy structure can dominate any smooth density profile that may be present. Furthermore, in Section~\ref{sec: J2051} it is suggested that the stochastic peaks in electron column density may determine the orbital phase and duration of the high-frequency ($>600$\,MHz) eclipses. Bearing this in mind, we note that much care must be taken when interpreting the relationship between eclipses and observing frequency, and suggest that higher-inclination systems may provide the best opportunity for theoretical modelling, in which the density distribution of the eclipse material sampled by the line of sight is likely to be less dominated by erratic clumpy structures. However, most eclipsing pulsars with reasonably constrained inclination angles appear to be far from being orientated edge-on \citep[Table~\ref{Table: psr_params};][]{bvr+13,cls+13,bkb+16}, thus, at the very least, attempts should consider the measured DM profile through the eclipse at the time of the observations.

\section{Conclusion}\label{sec: conclusion}
In this paper we present a direct comparison of the eclipses in two binary pulsars which have similar orbital parameters, but with companion star masses differing by an order of magnitude, namely, B1957 and J1816. Both systems show comparable eclipse durations, with DM variations of a similar order of magnitude at orbital phases near eclipse. The DM variations are shown to be highly asymmetric about the eclipses and fluctuate rapidly, indicative of a clumpy, swept back tail of material in both systems. The similar eclipse properties, regardless of the companion star type and mass, are discussed, and we consider that the eclipses may be independent of companion mass, or the effect of mass may be fortuitously cancelled out by other factors.\\
We estimated mass loss rates for the two systems, which lie in the range $10^{-13}$--$10^{-12}M_{\odot}$\,yr$^{-1}$, similar to those predicted for PSR J1810+1744 and PSR J2051$-$0827 in P18 and P19, again suggesting that the companions will not be fully evaporated on any reasonable timescale of evolution.\\
Simultaneous observations of the pulsed flux, and the continuum, unpulsed flux show that total removal of flux from the line of sight occurs during the main eclipse, which now appears to be common at low-frequencies. The observations are consistent with the expected effects of cyclotron-synchrotron absorption, or stimulated nonlinear scattering, as the primary causes of the eclipses.\\
For J1816 we present the first direct detection of an eclipse mechanism transferring between removal of flux, and smearing of pulsations in a single eclipse. We attribute this to scattering in the extended tail of material, and infer that the properties of this material must differ in B1957 as no such scattering is detected. Additionally, another potential difference between the two systems is presented, whereby measured pulse time-of-arrivals in B1957 appear to result from time-variable broadening of the pulses throughout the orbit, indicative of scattering or dispersion in what may be excess material extending around the orbit.\\
We further present measured full-eclipse, ingress and egress durations at a range of radio-frequencies for five spider pulsars, significantly increasing the volume of observational data available to constrain theoretical models. Out of the five pulsars, two had no previously measured frequency dependencies of the eclipse durations (J1816 and PSR J2051$-$0827), and for the remaining three we have increased the observed frequency coverage (PSR J1810+1744, B1957 and PSR J2215+5135).\\
For all of the pulsars there is a significant dependence of the eclipse durations on frequency, with low-frequencies corresponding to wider eclipses. We find that the commonly used power law functions generally give good fit to the data, but temporal variability of the eclipses can be a hindrance, thus (near-)simultaneous multi-frequency observations are required for reliable results. Our results strengthen previous suggestions that, at higher-frequencies, the eclipses become gradually more symmetric about inferior conjunction of the companion. We attribute this to a low-density tail of material trailing behind the companion in its orbit. However, in some cases (e.g. PSR J2051$-$0827) it appears that the stochastic nature of the electron density profile sampled by the line of sight can be the dominant effect in determining the location and duration of high-frequency eclipses.\\
The results offer a hint of correlation between the pulsar wind energy density at the distance of the companion star and the steepness of the power law. However, the relatively small amount of data and its limited reliability mean that this will need to be revisited when there is an improved knowledge of the incident pulsar wind energy and of the temporal variability of eclipse durations.

\section*{Acknowledgements}
EJP would like to thank Julian Donner, Caterina Tiburzi and Joris Verbiest for providing, and discussing, an ephemeris for PSR B1957+20 and GLOW station data for PSR J1810+1744. EJP would also like to thank Robert Main for discussions about PSR B1957+20, Jess Broderick for guidance with \textsc{prefactor}, and Stefan Os\l{}owski for observing with Parkes.\\
This paper is based (in part) on data obtained with the International LOFAR Telescope (ILT). LOFAR is the Low Frequency Array designed and constructed by ASTRON. It has observing, data processing, and data storage facilities in several countries, that are owned by various parties (each with their own funding sources), and that are collectively operated by the ILT foundation under a joint scientific policy. The ILT resources have benefitted from the following recent major funding sources: CNRS-INSU, Observatoire de Paris and Universit\'{e} d'Orl\'{e}ans, France; BMBF, MIWF-NRW, MPG, Germany; Science Foundation Ireland (SFI), Department of Business, Enterprise and Innovation (DBEI), Ireland; NWO, The Netherlands; The Science and Technology Facilities Council, UK; Ministry of Science and Higher Education, Poland.\\
The Parkes radio telescope is part of the Australia Telescope which is funded by the Commonwealth of Australia for operation as a National Facility managed by CSIRO.\\
Pulsar research at Jodrell Bank Centre for Astrophysics and Jodrell Bank Observatory is supported by a consolidated grant from the UK Science and Technology Facilities Council (STFC).\\
We thank the staff of the GMRT that made the observations possible. GMRT is run by the National Centre for Radio Astrophysics of the Tata Institute of Fundamental Research.\\
EJP acknowledges support from a UK Science and Technology Facilities Council studentship. RPB acknowledges support from the ERC under the European Union's Horizon 2020 research and innovation programme (grant agreement No. 715051; Spiders).




\bibliographystyle{mnras}
\bibliography{bibliography4} 






\bsp	
\label{lastpage}
\end{document}